\documentclass[useAMS,usenatbib]{mn2e}
\usepackage{graphicx}
\usepackage{amssymb}
\usepackage{lscape}
\usepackage{ulem}
\usepackage{txfonts}

\def\kms{km~s$^{-1}$}
\def\cm{cm$^{-2}$}
\def\lya{Ly$\alpha$}
\def\nhi{$N$(H\,{\sc i})}
\def\hi{H\,{\sc i}}
\def\si2{Si\,{\sc ii}}
\def\mg2{Mg\,{\sc ii}}
\def\fe2{Fe\,{\sc ii}}
\def\chr2{Cr\,{\sc ii}}
\def\al2{Al\,{\sc ii}}
\def\zn2{Zn\,{\sc ii}}
\def\mn2{Mn\,{\sc ii}}
\def\c2s{C\,{\sc ii}$^{\star}$}

\def\dind{$D$-index}

\title[Assessing dust selection bias at $0.7 < z< 1.6$]
{Assessing the dust selection bias  in quasar absorbers at $0.7 < z< 1.6$: 
Zn/Fe abundances in a radio-selected sample.\thanks{Based 
on observations made with ESO Telescopes at the Paranal Observatories 
under programmes 075.A-0158 and 076.A-0389.}}

\author[Ellison \& Lopez] {Sara L. Ellison$^1$ \& Sebastian Lopez$^2$.\\
$^1$Department of Physics and Astronomy, University of Victoria, Victoria, B.C., V8P 1A1, Canada\\
$^2$ Departamento de Astronom\'ia, Universidad de Chile, Casilla
     36-D, Santiago, Chile}

\begin{document}

\maketitle

\begin{abstract}
The Complete Optical and Radio Absorption Line System (CORALS) survey
has previously been used to demonstrate that the number density, gas
and metals content of $z>1.6$ damped Lyman alpha systems is not
significantly under-estimated in magnitude limited surveys.
In this paper, a sample of strong \mg2\ absorbers selected from 
the optically complete $0.7 < z < 1.6$ CORALS sample of Ellison et al. 
is used to assess the potential of dust bias at intermediate redshifts.
From echelle spectra of all CORALS absorbers with \mg2\ $\lambda$ 2796
and \fe2\ $\lambda$ 2600 rest equivalent widths $>$ 0.5 \AA\
in the redshift range $0.7 < z < 1.6$, we
determine column densities of Zn, Cr, Fe, Mn and Si.  The
range of dust-to-metals ratios and inferred number density of DLAs
from the $D$-index are consistent with
optical samples.  We also report the discovery of 4 new
absorbers in the echelle data in the redshift range $1.7 < z< 2.0$,
two of which are confirmed DLAs and one is a sub-DLA, whilst the
\lya\ line is not covered for the fourth.

\end{abstract}


\section{Introduction}

Galaxies contain dust.  Galaxies detected through their absorption
signature imprinted onto QSO spectra are no exception.  The presence
of dust in damped Lyman alpha (DLA) systems and other high column
density absorbers has been demonstrated ubiquitously through depletion
patterns of chemical elements (see the review by Wolfe et al. 2005)
and more occasionally through the direct detection of some extinction
feature, such as the 2175 \AA\ bump (Junkarrinen et al. 2004; Wang et
al 2004; Ellison et al. 2006).  In one case, the 9.7 $\mu$m silicate
feature has been detected (Kukarni et al. 2007).  Indirectly, the
presence of dust can be inferred through the ratios of refractory and
non-refractory elements to infer a gas-to-dust ratio, detection of
molecular absorption (e.g. Noterdaeme et al. 2008), abundance ratio
gradients in a DLA towards a lensed QSO (Lopez et al. 2005) and
diffuse interstellar bands (York et al. 2006; Ellison et al.  2008;
Lawton et al. 2008).  It is therefore perhaps unsurprising that for as
long as quasar absorption lines have been used to probe high redshift
galaxies (e.g. Wolfe et al. 1986), there have been concerns that dust
biases the results (e.g. Ostriker \& Heisler 1984).  Early work
indicated that a significant fraction of absorption-selected galaxies
might indeed be missed from surveys based on magnitude-limited optical
QSO samples (Fall, Pei \& McMahon 1989; Pei, Fall \& Bechtold 1991;
Fall \& Pei 1993).  More recent observations have tackled this problem
with either very large samples (Murphy \& Liske 2004; Vladilo,
Prochaska \& Wolfe 2008), or radio-selected QSO samples (Ellison et
al. 2001; Ellison, Hall \& Lira 2005; Jorgenson et al 2006).  Although
certain special classes of QSO absorbers have been shown to exhibit
substantial reddening, such as the Ca~II absorbers and very high
equivalent width \mg2\ absorbers (Wild, Hewett \& Pettini 2006; Menard
et al. 2008), in general reddening in the DLA absorbers is small.
Although a handful of individual reddening measurements have been
obtained (e.g. Vladilo et al. 2006) and a minority have quite high
inferred reddening values (e.g. Junkkarinen et al. 2004), the mean
E(B$-$V) in DLAs is low.  Vladilo et al. (2008) found E(B$-$V) $\sim$
0.005 mag in a large sample of DLAs selected from the magnitude
limited Sloan Digital Sky Survey.  Ellison et al. (2005) measured
E(B$-$V) $<$ 0.04 mag in an optically complete radio sample of DLAs.
It is also found that the number density and total neutral gas
fraction in radio-selected samples is in good agreement with large
optical samples (Jorgenson et al. 2006).  The typical metallicity of
radio-selected samples is also only marginally higher, by 0.2 dex,
than optical samples (Akerman et al. 2005).  Although the extinction
models presented by Vladilo \& P\'eroux (2005) suggest that
$\Omega_{\rm DLA}$ (the neutral gas mass fraction in DLAs) is
under-estimated by at least a factor of two by optical surveys, other
models have come up with much lower corrections (e.g. Trenti \&
Stiavelli 2006).  Most recently, Pontzen \& Pettini (2008) have used
Bayesian statistics to estimate that only 7\% of DLAs are missed from
magnitude limited DLA surveys with an under-estimate of $\Omega_{\rm
DLA}$ of 13\%.  They suggest that the larger corrections found by
Vladilo \& P\'eroux (2005) may be in part due to their smaller sample
of purely optically selected absorbers with measured abundances (that
are often biased by observers' preferential selection of high \nhi\
systems).

Although this substantial body of contemporary work has shown that
there is no convincing evidence for a substantial dust bias, these
results are largely limited to high redshifts, $z_{\rm abs} > 1.5$.
However, intermediate and low mass galaxies (log M$_{\star} \lesssim
10.5$ M$_{\odot}$) are still actively assembling their stellar mass at
this epoch with significant chemical enrichment occuring at lower
redshifts (e.g. Perez-Gonzalez et al. 2008; Panter et al.  2008).
Whilst one might reasonably expect
the chemical evolution of the general galaxy population to 
evolve monotonically with redshift, the reddening and extinction of background 
quasars is a function of the QSO and  absorber redshift, plus the 
extinction curve.  For example, in a previous CORALS paper by Ellison, Hall
\& Lira (2005), the goal was to search for a reddening signature, so that the
extinction curves and redshifts were accounted for in Monte Carlo simulations.
In the current paper, the chemical abundances are used as an 
indicator of whether
magnitude limited samples, on average, have distinct
abundance properties compared to optically complete surveys.
Given the smaller apertures of the space telescopes required to
confirm low redshift DLAs, quantifying the impact on chemical
abundances due to dust bias at $z<1.5$ is
particularly germane.  The only assessment of the impact of dust bias
at low redshifts was performed by Ellison et al. (2004).  Using an
optically complete radio-selected sample of quasars, Ellison et al.
(2004) determined that the number density of high equivalent width (EW) 
\mg2\ absorbers was in good agreement with optical samples.
30--50\% of these strong \mg2\ absorbers are expected to be DLAs
(Rao \& Turnshek 2000; Rao, Turnshek \& Nestor 2006), confirmation
which requires UV observations of the \lya\ line.  In the absence
of such data (and pending the installation of a UV spectrograph
on the Hubble Space Telescope, HST, during the next servicing
mission), we can nonetheless assess the dust depletion in the \mg2\
selected absorbers of Ellison et al. (2004).

In this paper, we present the results of new observations of 16 \mg2
-selected absorbers from the optically complete sample of Ellison et
al. (2004).  The new data consist of spectra of 15 QSOs obtained with
the UV and Visual Echelle Spectrograph (UVES) on the Very Large
Telescope (VLT), plus one archival spectrum obtained with the High
Resolution Echelle Spectrograph (HIRES) on Keck.  By comparing the
ratios of Zn and Fe column densities with observations from the
literature, it is shown that the dust depletion of these absorbers
is commensurate with optical samples.

\section{Sample selection}

In order to assess the effect of dust bias on absorption system
identification at $z<1.7$, Ellison et al. (2004) searched for \mg2\
absorbers in an optically complete sample of radio-loud QSOs.  Based
on the observation by Rao \& Turnshek (2000) and Rao et al.  (2006)
that absorbers with both high EWs of \fe2\ and \mg2\ have a high
chance of being DLAs, we selected all absorbers in the Ellison et
al. (2004) sample with EW(\fe2 $\lambda 2600) >$ 0.5 \AA\ and EW(\mg2
$\lambda 2796) >$ 0.5 \AA\ for high resolution follow-up.  There are
12 such absorbers that fulfill these criteria (see Table 3 of Ellison
et al. 2004)\footnote{From susbsequent observations, we found that the
high EW absorbers towards B0606$-$223 were not real.  Also, the
$z_{\rm abs}=0.7472$ absorber towards B1230$-$101 is now known to be a
superposition of many C~IV lines.  The absorbers in these two lines of
sight are therefore not included in our statistics.}.  In addition,
there are four more absorbers with EW(\mg2 $\lambda 2796) >$ 0.5 \AA\
but either no coverage of \fe2\ $\lambda 2600$, or with an upper limit
that is greater than 0.5 \AA\ reported in Ellison et al. (2004).  
In order to obtain a cleanly selected statistical sample, UVES spectra were
also obtained for these four sightlines.  Two of these four are 
found to have EW(\fe2 $\lambda 2600) >$
0.5 \AA\ from our UVES spectra.  In Table \ref{obs_table} we list the
16 absorbers, their redshifts\footnote{The absorption redshifts given
in Table \ref{obs_table} are the original values reported by Ellison
et al. (2004) for ease of reference with that work.  In Table
\ref{col_table} we give improved redshifts derived from the new
UVES data.} and \fe2\ and \mg2\ EWs as measured from
the original low resolution spectra (i.e. reproduced from Table 3 of
Ellison et al. 2004) or from the new UVES data (marked with a
$^{\star}$).  Since Rao et al. (2006) found that no DLA in their  
large (statistical) sample  of 197 \mg2\ systems at $z < 1.65$ has an EW(\mg2
$\lambda 2796) <$ 0.6 \AA, our echelle observations should include
all possible DLAs from the low redshift CORALS sample.  

\section{Observations and Data Reduction}

\begin{center}
\begin{table*}
\caption{Target List and Observing Journal}
\begin{tabular}{lcccccccccc}
\hline
QSO & B mag & $z_{\rm em}$ & $z_{\rm abs}$ &  EW(\mg2 $\lambda 2796)$ & EW(\fe2 $\lambda 2600$)  & Integration & S/N pix$^{-1}$ at & Observing\\
 & & & & (\AA) & (\AA) & Time (s) & \zn2 $\lambda$ 2026&  Period \\
\hline
B0039$-$407 & 19.7 & 2.478  & 0.8483   & 2.35 $\pm$ 0.15 & 2.09 $\pm$ 0.22       & 16,500  & 7 & P75\\
B0122$-$005 & 18.5 & 2.280  &  0.9949  & 1.56 $\pm$ 2.08 & 0.56 $\pm$ 0.38        & 13,200  & 35 & P75\\
B0227$-$369 & 19.6 & 2.115  &  1.0289  & 0.59 $\pm$ 0.18 & 0.61 $\pm$ 0.16        & 12,800  & 7 &  P76\\
B0240$-$060 & 18.7 & 1.805  &  0.5810  &1.44 $\pm$ 0.08 & 0.69 $\pm$ 0.05$^{\star}$         & 6000  &  ... & P76\\
             & &       &  0.7550  &1.65 $\pm$ 0.04 & 1.25 $\pm$ 0.04         & & 7  &  \\
B0244$-$128 & 18.4 & 2.201  &  0.8282  &1.77 $\pm$ 0.09 & 1.23 $\pm$ 0.09        & 6000 & 26&  P76\\
B0458$-$020 & 19.0 & 2.286  &  1.5606  &0.94 $\pm$ 0.08 & 0.75 $\pm$ 0.08         &  28,800     & 25 &  Feb 1995 \\
B0919$-$260 & 19.0 & 2.300  &  0.7048  &0.81 $\pm$ 0.08 & $<0.36^{\star}$        & 13,200& 9 &P75\\
B1005$-$333 & 18.0 & 1.837  &  1.3734  &0.93 $\pm$ 0.10 & 0.84 $\pm$ 0.11        & 6600 & 8&  P75 \\
B1256$-$177 & 21.4 & 2.263  &  0.9399  &2.96 $\pm$ 0.04 & 2.08 $\pm$ 0.04        & 13,200  &  7 & P75\\
B1318$-$263 & 21.3 & 2.027  &  1.1080  & 1.38 $\pm$ 0.08 & 0.61 $\pm$ 0.07        & 23,100  & 3&  P75\\
B1324$-$047 & 19.8 & 1.882  &  0.7850  &2.58 $\pm$ 0.12 & 1.77 $\pm$ 0.11        & 19,800  & 7 &  P75\\
B1402$-$012 & 18.0 & 2.518  &  0.8901  &1.21 $\pm$ 0.04 & 0.99 $\pm$ 0.04        & 5300&  10 &P75\\
B1412$-$096 & 17.4 & 2.001  &  1.3464  &0.66 $\pm$ 0.11 & 0.26$\pm$0.2$^{\star}$       & 3300  & 10& P75\\
B2149$-$307 & 18.4 & 2.345  &  1.0904  &1.45 $\pm$ 0.04 & 0.78 $\pm$ 0.04        & 5200 & 23& P75\\
B2245$-$128 & 18.3 & 1.892  &  0.5869  &1.28 $\pm$ 0.08 & 0.51$\pm$0.03$^{\star}$       & 5200  & ... & P75 \\ \hline
\multicolumn{9}{l}{\textit{Notes}: Magnitudes taken from Ellison et al. (2004).
Equivalent widths are from the original low resolution
data, except where marked with an }\\
\multicolumn{9}{l}{ asterisk ($^{\star}$) where they are measured from
the UVES spectra. Absorbers with EW(\fe2 $\lambda 2600$) $<$ 0.5 \AA\ 
are excluded from the statistical sample.}
\end{tabular}\label{obs_table}
\end{table*}
\end{center}

B0458$-$020 was observed with HIRES in February 1995 and the reduced
spectrum, covering a wavelength range of 3900-6300 \AA\ was provided
to us by J. X. Prochaska.  All other data have been newly acquired
with the UVES spectrograph on the VLT during ESO observing Periods 75
and 76 (April 2005 -- March 2006).  In all cases we used the 390+564
setting which gave wavelength coverage from approximately 3300--4500
\AA\ and 4600--6650 \AA.  The exception to this was B1402$-$012 which
was observed with the 437-760 setting, giving coverage from 3750--5000
\AA\ and 5700--9450 \AA.  In Table \ref{obs_table} we list the
exposure times for all UVES and HIRES observations.  The UVES spectra
were extracted and reduced using the UVES pipeline. The reduced
extractions of each exposure were converted to a common
vacuum-heliocentric scale, and then stacked weighting by the inverse
of the flux variances.

\section{Column densities and relative abundances}

The average spectral S/N ratio of our sample is lower than is typically
obtained for abundance determinations.  This is a result of the fact
that an optically complete sample, such as CORALS from which these
targets are drawn, will include many fainter QSOs than a magnitude
limited sample.  The lower S/N ratios hamper detailed Voigt profile
decomposition, so that we instead determine total column densities
using the apparent optical depth method (AODM, e.g. Savage \& Sembach
1991).  This technique is a widely used alternative to Voigt fitting
of QSO absorption lines  (e.g. Prochaska et al. 2001).  Although the
AODM is potentially susceptible
to `hidden saturation' DLA column densities derived with this method generally
give good agreement with other techniques.  Although Fox,
Savage \& Wakker (2005) showed that the AODM may over-estimate column
densities by $\sim$ 10\% in low S/N data, the relative abundances
derived here should be unaffected by systematic offsets.  We derive
column densities (or limits) for \fe2, \zn2, \chr2 and \mn2\ in the
majority of our absorbers.  \si2\ can be determined for only a small
number of systems due to the low redshift range of the absorbers.
When more than one transition is observed for a given species (as is
often the case for \fe2), we compare
the column densities derived from each to test for possible saturation
(i.e. the highest $f$-value line yields a column density which
is significantly lower than the other transitions).  A line that is suspected
to be saturated is discarded (this process is in addition to the rejection
of lines that are clearly saturated with zero residual flux in their line
cores).  In the
absence of suspected saturation, the average of the column
densities from multple transitions is adopted.
Absorption line profiles of selected transitions for our QSO sample
are shown in Figures \ref{B0039} -- \ref{B2245}.  
We use the atomic data tabulated by
Morton (2003) which is reproduced in Table \ref{atom_data} for ease of
reference.  The measured column densities are given in Table
\ref{col_table}, as well as improved absorption redshifts derived
from the UVES data.  We adopt the usual assumption that the singly
ionized species studied here approximate the total column density of a
given element.  Although there are likely to be ionization corrections
for the lower \nhi\ absorbers, we do not have the necessary data to
estimate these corrections.  Dessauges-Zavadsky (2003) have previously
estimated that ionization corrections are likely to be on the order of
0.1 -- 0.2 dex.  However, such corrections are sensitive to the input
ionizing spectrum and so the corrections could be
significantly larger (Howk \& Sembach 1999; Meiring et al. 2007).  
Corrections for zinc are further
complicated by large uncertainties in the atomic data (Howk \& Sembach 1999;
Dessauges-Zavadsky et al. 2003).
Although ionization corrections may therefore lead to errors in our
derived column densities, this same problem should be present in the
comparison data taken from the literature (at a given \nhi).  
A differential comparison
between the abundance ratios in the CORALS absorbers and the optical
samples taken from the literature is therefore still feasible.
Observed column densities are used to derive relative abundances on
the solar scale (see Table \ref{solar_data}) and are given in Table
\ref{abund_table}.  For completeness, Table \ref{abund_table}
includes the abundances of the two absorbers (towards B0919$-$260
and B1412$-$096) subsequently found to have EW(\fe2\ $\lambda$ 2600)
$<$ 0.5 \AA, although these absorbers are not included in the statistical
comparisons discussed in the next section.

\begin{center}
\begin{table*}
\caption{\label{col_table}Measured column densities and 3 $\sigma$ limits for the CORALS MgII absorbers} 
\begin{tabular}{lcccccc}
\hline
QSO & $z_{\rm abs}$ & Log N(\fe2) & Log N(\chr2) & Log N(\zn2) & Log N(\si2) & Log N(\mn2) \\ \hline
B0039$-$407  &  0.84885  & $>$15.0 & 13.3$\pm$0.2 & ...& ...  & 13.2$\pm$0.1\\ 
B0122$-$005  &  0.99430  & 14.08$\pm$0.07 & $<$12.1  & 11.8$\pm$0.2  & ... & 11.9$\pm$0.2 \\ 
B0227$-$369  &  1.02900  & $>$14.6 & $<$12.6  & 12.5$\pm$0.2 & 15.1$\pm$0.2 & 12.5$\pm$0.2 \\ 
B0240$-$060  &  0.58103  & 14.1$\pm$0.2 & ... &...  &...& $<$11.9 \\ 
B0240$-$060  &  0.75468  & 14.69$\pm$0.08 & $<$12.7 & $<$12.1  &... & ... \\ 
B0244$-$128  &  0.82850  & 15.0$\pm$0.2 & 13.3$\pm$0.2 & 12.8$\pm$0.1  &... & 12.9$\pm$0.2 \\ 
B0458$-$020  &  1.56055 & 14.92$\pm$0.05 & 13.09$\pm$0.05 & 12.48$\pm$0.05 & 15.3$\pm$0.1 &...\\ 
B0919$-$260  &  0.70527 & 13.61$\pm$0.07 & $<$12.4 & $<$11.6  & ... & $<$11.7 \\
B1005$-$333  &  1.37381  & 14.6$\pm$0.2 & $<$ 12.8  & 12.3 $\pm$ 0.3 & ...&$<$11.9 \\ 
B1256$-$177  &  0.93495  & $>$14.7 & $<$12.7 &12.3$\pm$0.2  & $<$14.8 & ... \\ 
B1318$-$263  &  1.10407  & 14.2$\pm$0.1 & $<$12.8 & $<$12.4 & ... & 12.4$\pm$0.3\\ 
B1324$-$047  &  0.78472  & 15.3$\pm$0.2 &13.2$\pm$0.2  &12.9$\pm$0.3  &... & 13.1$\pm$0.2 \\ 
B1402$-$012  &  0.88978  & 14.6$\pm0.1$ & $<$12.7 & 12.5$\pm$0.2 & ... &12.5$\pm$0.2 \\
B1412$-$096  &  1.34652  & 13.5$\pm$0.2 & $<$12.7 & $<$11.9 & $<$14.5 & $<$12.1\\ 
B2149$-$307  &  1.09074  & 14.25$\pm$0.09 &  $<$12.2 & $<$11.5 & 14.9$\pm$0.2 & $<$11.6 \\ 
B2245$-$128  &  0.58700  & 14.2$\pm$0.1 & ...  & ...  & ... & $<$11.8 \\ 
 \hline 
\multicolumn{7}{l}{\textit{Note}: Limits are 3$\sigma$ and assume a
line width of 7 \kms.}
\end{tabular}
\end{table*}
\end{center}

\begin{center}
\begin{table}
\caption{\label{solar_data}Solar meteoritic adundances from Lodders (2003)} 
\begin{tabular}{cc}
\hline
Element, X & [X/H]$_{\odot}$ \\ \hline
Fe & $-4.53$ \\
Si & $-4.46$ \\
Zn & $-7.37$ \\
Cr & $-6.35$ \\
Mn & $-6.53$ \\
 \hline 
\end{tabular}
\end{table}
\end{center}

\begin{center}
\begin{table}
\caption{\label{abund_table}Relative Abundances on the Solar Scale} 
\begin{tabular}{lcrrrr}
\hline
QSO &  $z_{\rm abs}$ & [Cr/Fe] & [Zn/Fe] & [Si/Fe] & [Mn/Fe] \\ \hline
B0039$-$407  &  0.84885  & $<+$0.1 &...  & ... & $<+$0.2\\ 
B0122$-$005  &  0.99430  & $<-0.2$ & +0.6  &... &$-$0.2\\ 
B0227$-$369  &  1.02900  & $<-0.2$ & $<+0.7$  & $<+0.4$ & $<-0.1$\\ 
B0240$-$060  &  0.58103  &...  & ... &... &$<-0.2$\\ 
B0240$-$060  &  0.75468  &$<-0.2$  & $<0.3$ &... &...\\ 
B0244$-$128  &  0.82850  & +0.1 & +0.6  &...& $-0.1$\\ 
B0458$-$020  &  1.56055  & $-0.01$ & +0.40 & +0.31 & ... \\ 
B0919$-$260  &  0.70527  & $<+0.6$ & $<+0.8$ &...& $<+0.2$\\ 
B1005$-$333  &  1.37381  & $<0$ & +0.5  & ... & $<-0.7$\\ 
B1256$-$177  &  0.93495  & $<-0.2$ &$<+0.4$  & $<+0.0$ & ...\\ 
B1318$-$263  &  1.10407  & $<+0.4$ & $<+1.0$ &...& +0.2\\ 
B1324$-$047  &  0.78472  & $-0.3$ & +0.4  &... & $-0.2$\\ 
B1402$-$012  &  0.88978  & $<-0.1$ & +0.7 & ...& $-0.1$\\ 
B1412$-$096  &  1.34652  & $<+1.0$ & $<+1.2$  & $<+0.9$ & $<+0.6$\\ 
B2149$-$307  &  1.09074  & $<-0.2$ & $<+0.1$  & +0.6 & $<-0.7$\\ 
B2245$-$128  &  0.58700  & ...  &  ...& ...& $<-0.4$\\ 
 \hline 
\end{tabular}
\end{table}
\end{center}

\section{Results}

\begin{figure}
\centerline{\rotatebox{270}{\resizebox{6cm}{!}
{\includegraphics{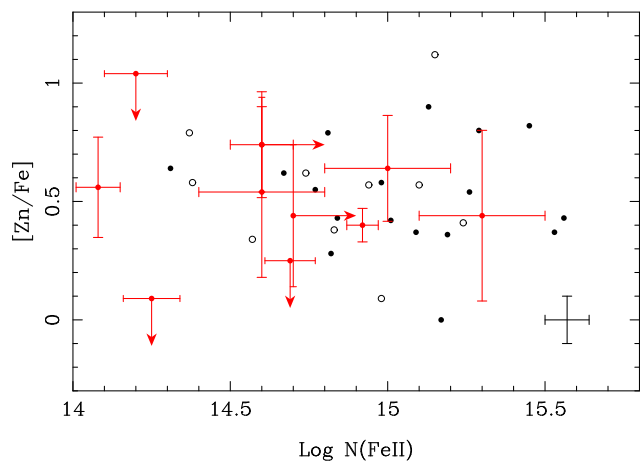}}}}
\caption{Comparison of [Zn/Fe] as a function of log N(\fe2) of
literature $0.6 < z_{\rm abs} < 1.7$ DLAs (solid black points) and
sub-DLAs (open black points) with the \mg2 -selected CORALS absorber
(points with error bars) sample. For clarity, error bars for the
literature data are not shown, but the typical error bar is shown in
the lower right corner of the plot.\label{znfe} }
\end{figure}

Akerman et al. (2005) used high resolution spectra of the high
redshift CORALS DLA sample to demonstrate that the metallicity (as
inferred from the Zn abundance) was only marginally higher (by 0.2
dex) than the literature value at $z \sim 2.5$.  Combined with the
observation that the [Zn/Cr] ratios were also consistent with previous
magnitude limited samples, Akerman et al. (2005) concluded that, at
high redshift, dust bias does not lead to a sample that is
artificially dominated by metal-poor, low dust absorbers.  In the
absence of neutral hydrogen column densities (only available with a
space telescope), we can only use the relative abundances to
investigate the possibility of dust bias.  A second potential
disadvantage of not having \nhi\ is that a significant fraction of
absorbers in our sample are likely to be sub-DLAs with column
densities log \nhi\ $<$ 20.3 \cm.  Fortunately, Dessauges-Zavadsky et
al.  (in preparation, see also Dessauges-Zavadsky, Ellison \& Murphy
2009) have used the largest sample\footnote{The sample consists of
approximately 200 DLAs (log \nhi\ $\ge$ 20.3) and 100 sub-DLAs (19.0 $<$
log \nhi\ $<$ 20.3) with absorption redshifts ranging from 0.01 to
4.75.} of intermediate redshift sub-DLAs yet compiled to demonstrate
that the [Zn/Fe] ratios in sub-DLAs are statistically
indistinguishable from the DLAs.  It is therefore possible to compare
the [Zn/Fe] relative abundances in our \mg2\ selected sample with a
literature compilation of DLAs and sub-DLAs.

The range of Zn and Fe column densities in our sample is commensurate
with optical samples in the literature.    In
Figure \ref{znfe} we show the relative abundances of Zn and Fe of the
\mg2 -selected CORALS absorbers with sub-DLAs and DLAs taken from the
compilation of Dessauges-Zavadsky et al. (in preparation) whose
absorption redshifts are in the range $0.6 < z_{\rm abs} < 1.7$.  If
the literature sample is systematically biased against absorbers with
high dust depletion, we expect that their [Zn/Fe] ratios will, in
general, be lower (at a given N(\fe2)).  Visually, the figure shows
that the CORALS sample has [Zn/Fe] ratios consistent with the
literature and the two samples have identical mean values of
[Zn/Fe]=0.55 (detections only).  This result is confirmed
statistically: a 2D Kolmogorov-Smirnov (KS) test shows that there is
no significant difference in the N(\fe2)--[Zn/Fe] distribution of the
CORALS and literature samples\footnote{This is true regardless of
whether limits are included or whether only detections are used in the
KS test.}.  Although the samples are not selected identically (the
CORALS absorbers are selected on \fe2\ and \mg2\ EWs, whilst the
literature sample is selected on both on metal line EWs and \nhi),
Figure \ref{znfe} nonetheless demonstrates that the CORALS absorbers
do not have particularly high dust-to-metals ratios.  The one
intriguing exception to this general statement is the absorber towards
B1318$-$263.  Although Zn is not detected for this system, the [Mn/Fe]
is super-solar by +0.2 dex, which is unusual amongst DLAs
(e.g. Pettini et al. 2000; Dessauges-Zavadsky et al. 2002).
Super-solar [Mn/Fe] is seen in the Galactic ISM in cool disk clouds
(e.g. Savage \& Sembach 1996) and is usually interpreted as a
signature of depletion.  Only one DLA in the compilation of
Dessauges-Zavadsky et al. (2002) has [Mn/Fe] $>$ +0.1.  The DLA
towards B1318$-$263 may therefore be amongst the most depleted.
However, even in this case the [Zn/Fe] is constrained to be $<$ +1.0,
a value which is high amongst DLAs (see Figure \ref{znfe}), but still
low compared to Galactic disk sightlines (e.g. Savage \& Sembach
1996).

\section{Newly discovered absorbers in the UVES spectra}

\begin{figure}
\centerline{\rotatebox{0}{\resizebox{9cm}{!}
{\includegraphics{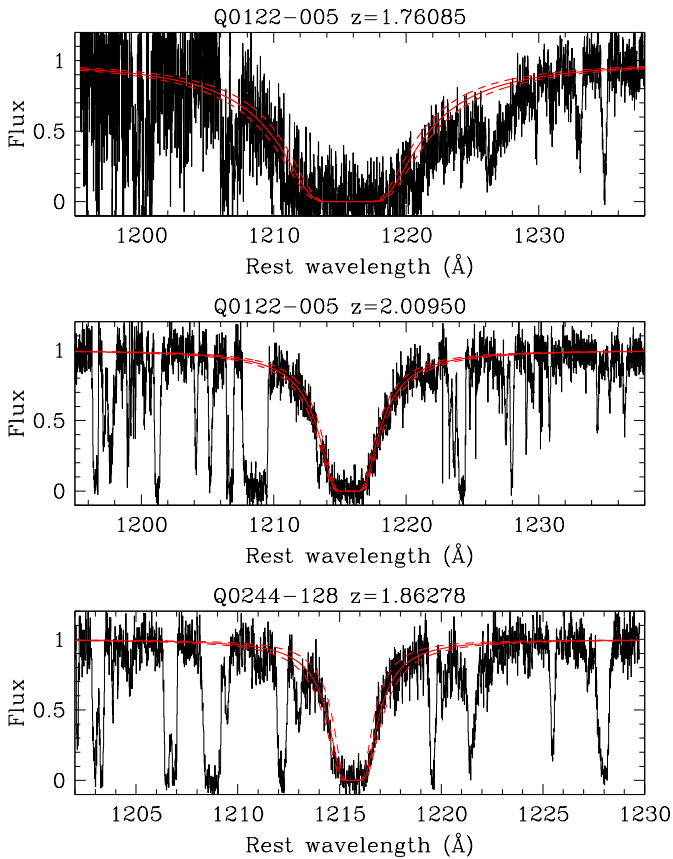}}}}
\caption{\label{HI_fits} Fits to the \lya\ profiles for the three
newly discovered \hi\ absorbers in our UVES data.  Column densities
and uncertainties (dashed profiles) are given in Table \ref{col_table_new}. }
\end{figure}

The UVES data obtained for this study have also led to the discovery
of four new absorbers which we present here for completeness.  We have
determined \nhi\ column densities through fully damped Voigt-profile
fits of the \lya\ lines, which are shown in Figure \ref{HI_fits}.
Selected metal lines of the all four absorbers are shown in Figures
\ref {B0122_1} -- \ref{B2149_2}.  The new absorbers consist of 1 DLA,
2 sub-DLAs and one absorber whose \lya\ transition is not covered.
This latter system is likely to be a DLA or high \nhi\ sub-DLA based
on its strong metal lines.  The column densities were derived through
the AODM method and are presented in Table \ref{col_table_new}.

\begin{center}
\begin{table*}
\caption{\label{col_table_new}Measured column densities for the newly discovered absorbers in the UVES spectra} 
\begin{tabular}{lcccccc}
\hline
QSO & $z_{\rm abs}$ & N(\hi) ($10^{19}$ \cm) & Log N(\fe2) & Log N(\chr2) & Log N(\zn2) & Log N(\si2)  \\ \hline
B0122$-$005  &  1.76085  & 60$\pm$10 &  15.1$\pm$0.1 & 13.4$\pm$0.2 & ... & 15.42$\pm$0.08 \\
B0122$-$005  &  2.00950  & 11$\pm$2 & 13.69$\pm$0.07 & $<$12.1 & $<$11.4 & 13.67$\pm$0.05  \\ 
B0244$-$128  &  1.86278  & 3$\pm$1 & $<$13.9 & 12.2$\pm$0.3 & $<$11.5 & 14.2$\pm$0.2 \\
B2149$-$307  &  1.70085  & ... & 15.0$\pm$0.1& 13.2$\pm0.2$ & 12.6$\pm$0.1  & 15.4$\pm$0.1 \\ 
 \hline 
\end{tabular}
\end{table*}
\end{center}

\section{Summary and discussion}

Akerman et al. (2005) showed that the dust-to-metals ratio of CORALS
DLAs was consistent with DLAs from magnitude limited literature
samples.  It was argued that this was evidence against a particularly dusty
population of DLAs at $1.7 < z_{\rm abs} < 3.5$.  In this paper, we
have extended that conclusion down to lower redshifts, where galaxies
may be more chemically evolved.  We have compiled an optically complete
sample of absorbers with strict selection criteria of EW(\fe2\ $\lambda$
2600, \mg2\ $\lambda 2796$) $>$ 0.5 \AA.  In our sample of 14 $0.6 < z_{\rm abs}
< 1.7$ \mg2 -selected absorbers fulfilling these EW criteria, 
we can determine values or limits for
[Zn/Fe] for 11 absorbers.  The relative abundances are commensurate
with the distribution in the literature for DLAs and sub-DLAs.  This
result adds to the growing body of observational evidence that,
although dust is present in absorption selected galaxies, the
observational bias imposed by this dust does not strongly affect
the statistical properties of the DLA population.

The obvious next step from the current work is to determine \nhi\ for
the CORALS absorbers presented here.  \hi\ column density
measurements would allow us to a) confirm which of the absorbers are
truly DLAs, b) determine absolute abundances and c) derive spin
temperatures for the subset of absorbers with 21cm observations
presented in Kanekar et al. (2009).  Out of the 14 \mg2\ absorbers
in our EW-selected sample, Kanekar et al. (2009) have obtained
21 cm absorption spectra for 11.  Nine absorbers have both measurements
(or limits) of the Zn column density and 21 cm optical depth.
Based on a dataset of 25 points over all redshifts, Kanekar 
et al. (in preparation) show that metallicity may be anti-correlated 
with spin temperature.  However, since it has been suggested that
the 21cm detection rate and spin temperature of DLAs may evolve 
with redshift (e.g. Kanekar et al. 2003; Kanekar et al. 2009),
with only one known DLA at $z > 1.7$ exhibiting a spin temperature
below 500 K (York et al. 2007), it would be desirable to consider
the relation of spin temperature with metallicity as a function
of redshift.  Only 9 absorbers currently have measurements of
both Zn and 21 cm optical depth at $z < 1.7$.  Obtaining \lya\
observations of the nine systems in the current sample with measurements
of both N(\zn2) and $\tau_{\rm 21cm}$ would therefore double
the low redshift sample in which this relation could be investigated.
 
\hi\ observations (at the present time) require HST
spectroscopy and are therefore pivotal on the successful installation
of the Cosmic Origins Spectrograph (COS) and/or the repair/upgrade of
the Space Telescope Imaging Spectrograph (STIS) scheduled for
Servicing Mission 4.  The design of such 
observations would prioritise likely DLAs over
sub-DLAs in order to make the direct comparison of metallicities in
magnitude limited and optically complete samples.  Although
the current work has shown that dust-to-gas ratios in the low redshift
CORALS sample are typical, a comparison of metallicities
(as done by Akerman et al. 2005 for the high redshift CORALS sample)
is required in order to assess the fraction of missing metals.
The high resolution spectra presented here can
be used to guide those observations, by assessing which of the \mg2
-selected absorbers are unlikely to be DLAs.  Ellison (2006)
investigated how the combination of EW and kinematic information from
high resolution spectra could be used to screen for DLAs.  Ellison (2006)
introduced the \dind\ which is a measure of saturation over the velocity
profile over the \mg2\ $\lambda$2796 line with DLAs tending to have
higher values than sub-DLAs, as measured in high resolution data.
Specifically, Ellison (2006) defined 

\begin{equation}\label{d_eqn}
D = \frac{EW (\AA)}{\Delta v (km/s)} \times 1000,
\end{equation}

where the EW and velocity spread of the \mg2\
$\lambda$2796 line are measured over 90\% of the EW
of the main absorption component.   In a more recent
study, Ellison et al. (2008) have shown that including the
\fe2\ column density slightly improved the discriminating
power of the D-index and defined

\begin{equation}\label{dfe_eqn}
D_{\rm Fe} = \frac{EW (\AA)}{\Delta v (km/s)} \times \frac{\log N(FeII)}{15} \times 1000
\end{equation}

Ellison et al. (2008) found that $\sim$ 54 and 57\% of absorbers with
$D > 7.2$ and $D_{\rm Fe}> 7.0$ respectively are DLAs.

We calculate the D-index for the absorbers in our sample using the
UVES data.  For B0458$-$020, the HIRES spectra did not cover the \mg2
$\lambda$ 2796 line, but a UVES spectrum was available in the archive
that did possess the requisite wavelength coverage.  For 2 of
our absorbers (B0227$-$369 and B1402$-$012), the \mg2\ line lies in
the UVES chip/inter-arm gaps, so the D-index could not be calculated.

In Table \ref{d_table} we present the values of $D_{\rm Fe}$ for our
\mg2\ sample.  Again, for completeness, we include the indices for the
two absorbers with EW(\fe2\ $\lambda$) $<$ 0.5 \AA, but do not
consider them in our statistics.  From the statistical sample of 14
absorbers, nine have $D_{\rm Fe}> 7.0$, three have $D_{\rm Fe}< 7.0$
and two have insufficient information to judge.  The three low \dind\
absorbers are unlikely to be DLAs (see Figure 7 of Ellison et
al. 2008).  However, only five of the nine high \dind\ absorbers are
statistically likely to be DLAs, based on the success rate of the
D-index.  We therefore statistically expect 5/12 = 42\% to actually be
DLAs.  This is in good agreement with the fraction of DLAs expected
from \mg2\ and \fe2\ EWs alone (which formed the initial selection of
the sample), 36$\pm$6\% as found by Rao et al. (2006).  This DLA
fraction can be scaled up to the full statistical sample of 14
absorbers and combined with the redshift paths in Ellison et
al. (2004) to determine the number density of DLAs in the intermediate
redshift CORALS sample.  Using errors appropriate for small samples
(Gehrels 1996), we determine $n(z) = 0.11^{+0.07}_{-0.04}$ for a mean
absorption redshift of $\langle z_{\rm abs} \rangle =
$1.0\footnote{The derived n(z) for DLAs based on the D-index is
slightly lower than the value found by Ellison et al. (2004) (n(z) =
0.16$^{+0.08}_{-0.06}$) for the CORALS \mg2\ absorbers based on EW
statistics.  The main difference is due to updated results from Rao
et al. (2006) regarding the likelihood that a strong \mg2\ absorber is
a DLA, a probability that has dropped from 50\% to 35\% since Rao \&
Turnshek (2000).  Moreover, Ellison et al. (2004) used 5$\sigma$
significance limits, whereas we have adopted 3$\sigma$.  Taking into
account these two differences, the DLA n(z) derived purely from EW
statistics for CORALS is 0.13$^{+0.07}_{-0.05}$, in good agreement
with the incidence rate derived from D-index statistics.}.  It should
be noted that the original intermediate redshift CORALS sample
presented in Ellison et al. (2004) omitted two QSOs due to their faint
magnitudes and limited observing time.  The detection of an absorber
with \mg2, \fe2\ EW $>$ 0.5 \AA\ and a \dind\ $>$ 7.0 in either one or
both of these sightlines would increase the inferred number density to
n(z) = 0.12 and 0.14 respectively.  The n(z) is therefore in excellent
agreement with the value of $n(z) = 0.120\pm0.025$ at $\langle z_{\rm
abs} \rangle = $1.2 derived by Rao et al. (2006) from their HST
survey.  Taken together with the consistency of Zn/Fe ratios between
the CORALS DLAs and the literature, the concordant DLA number
densities of CORALS and the HST survey (despite the relatively bright
magnitude limit of the latter), implies that optically selected
samples of intermediate redshift DLAs are not missing a significant
population of absorbers, and that their dust depletion properties are
representative of the full population of DLAs.

\begin{center}
\begin{table}
\caption{\label{d_table}D-indices for the CORALS MgII absorbers} 
\begin{tabular}{lcc}
\hline
QSO & $z_{\rm abs}$ &      D$_{\rm Fe}$ \\ \hline
B0039$-$407   &  0.84885  &  $\ge 8.94$         \\
B0122$-$005   &  0.99430  &  5.15$\pm$0.19  \\
B0227$-$369   &  1.02900  & ...       \\
B0240$-$060   &  0.58103  &  7.49$\pm$0.23 \\
B0240$-$060   &  0.75468  &  5.43$\pm$0.43 \\
B0244$-$128   &  0.82850  &  9.11$\pm$0.12 \\
B0458$-$020   &  1.56055  &  6.75$\pm$3.39 \\
B0919$-$260   &  0.70527  &  5.60$\pm$0.22 \\
B1005$-$333   &  1.37381  &  8.74$\pm$0.58 \\
B1256$-$177   &  0.93495  &  $\ge 8.30$      \\
B1318$-$263   &  1.10407  &  7.25$\pm$0.85\\
B1324$-$047   &  0.78472  &  8.51$\pm$0.31\\
B1402$-$012   &  0.88978  & ...   \\
B1412$-$096   &  1.34652  &  6.59$\pm$1.00\\
B2149$-$307   &  1.09074  &  7.91$\pm$0.25\\
B2245$-$128   &  0.58700  &  7.03$\pm$0.43\\
 \hline 
\end{tabular}
\end{table}
\end{center}

\section*{Acknowledgments} 

SLE is supported by an NSERC Discovery Grant and SL was partly
supported by the Chilean {\sl Centro de Astrof\'\i sica} FONDAP
No. 15010003, and by FONDECYT grant N$^{\rm o}1060823$..  We are
grateful to J. X. Prochaska for providing the HIRES spectrum of
B0458$-$020 and to Nissim Kanekar for useful comments on an
earlier draft of this paper.

\begin{appendix}
  
\section{Absorption line figures}

\begin{figure}
\centerline{\rotatebox{0}{\resizebox{7cm}{!}
{\includegraphics{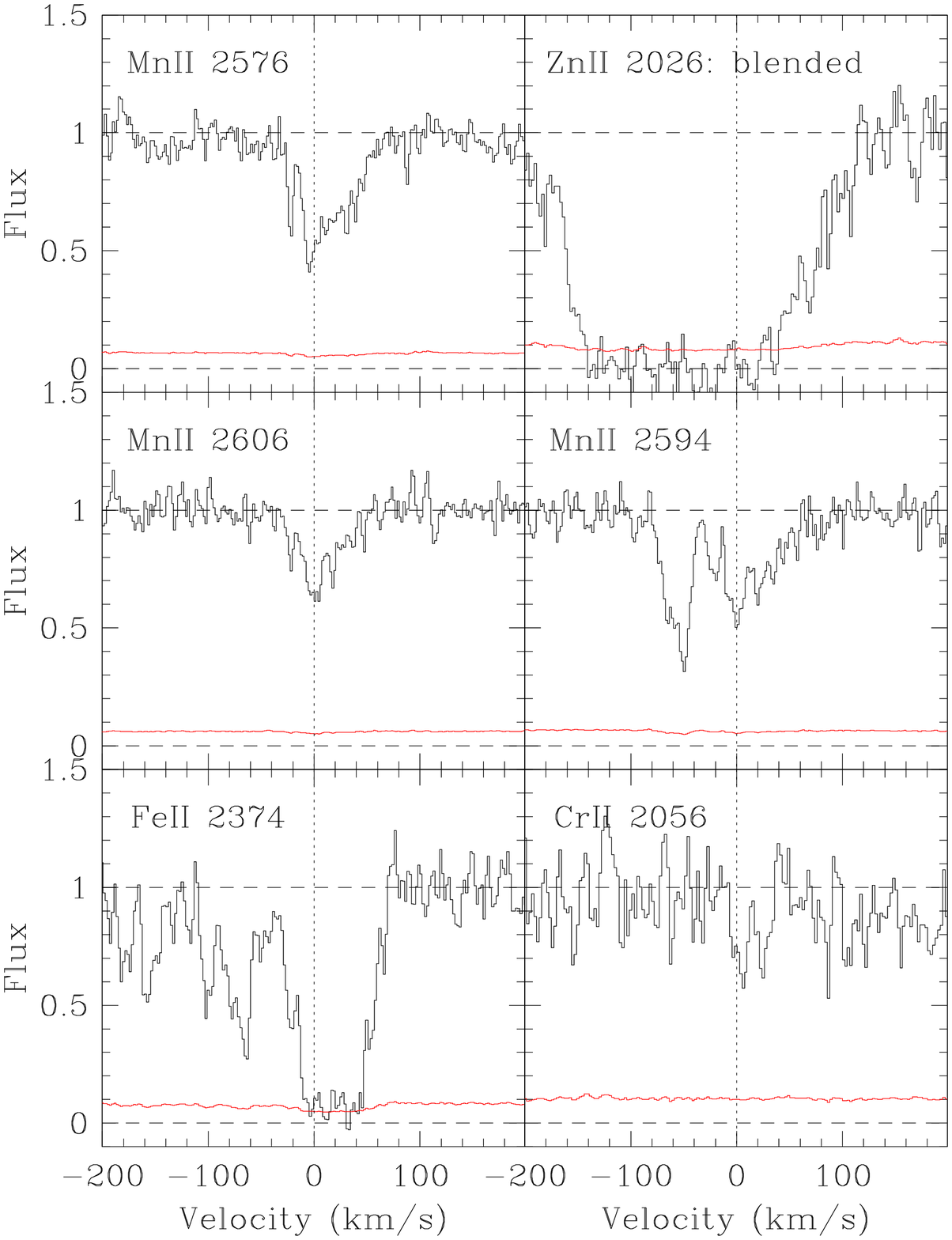}}}}
\caption{\label{B0039}Metal lines towards B0039$-$407 on a velocity scale 
relative to $z=0.84885$  }
\end{figure}

\begin{figure}
\centerline{\rotatebox{0}{\resizebox{7cm}{!}
{\includegraphics{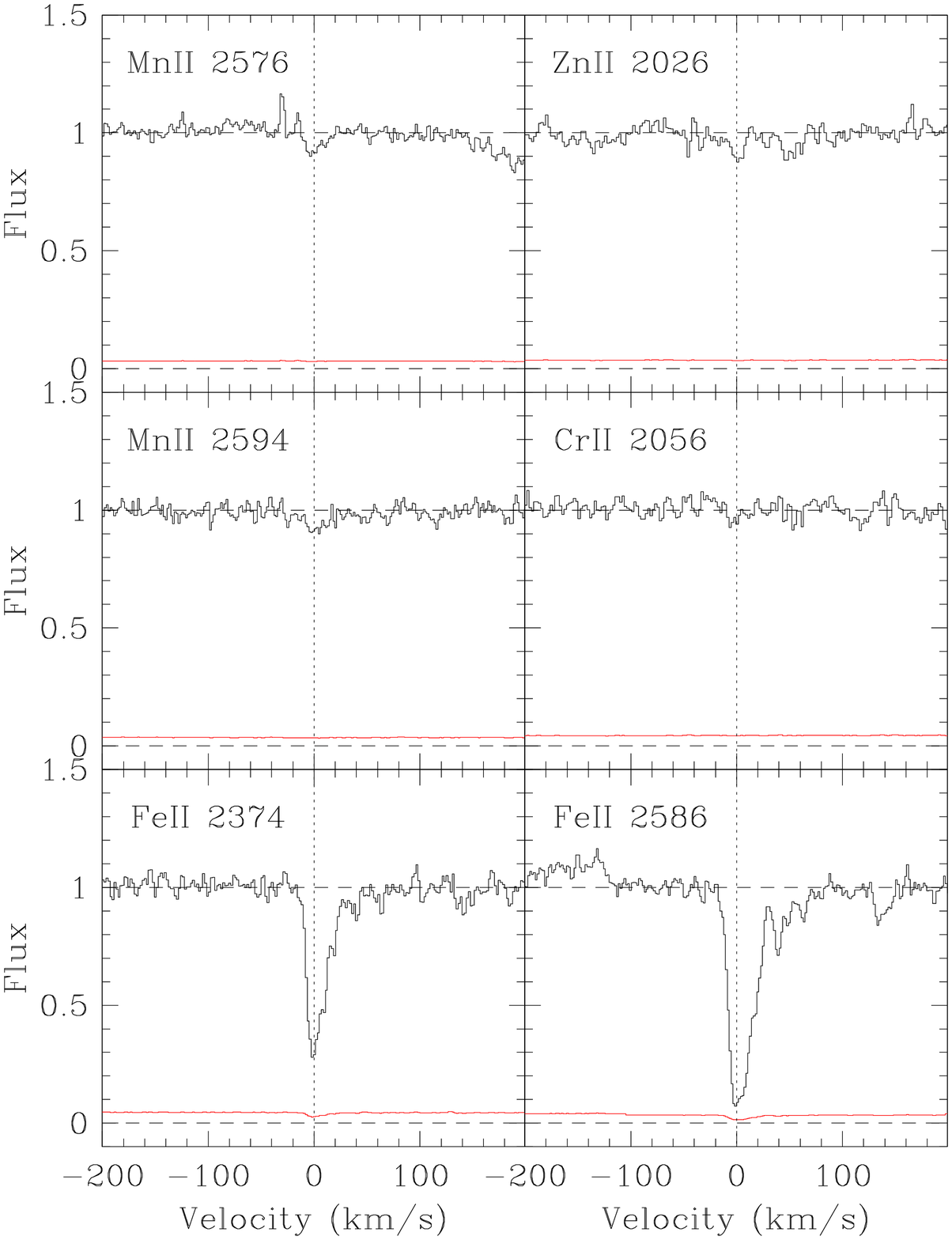}}}}
\caption{\label{B0122}Metal lines towards B0122$-$005 on a velocity scale 
relative to $z=0.99430$  }
\end{figure}

\begin{figure}
\centerline{\rotatebox{0}{\resizebox{7cm}{!}
{\includegraphics{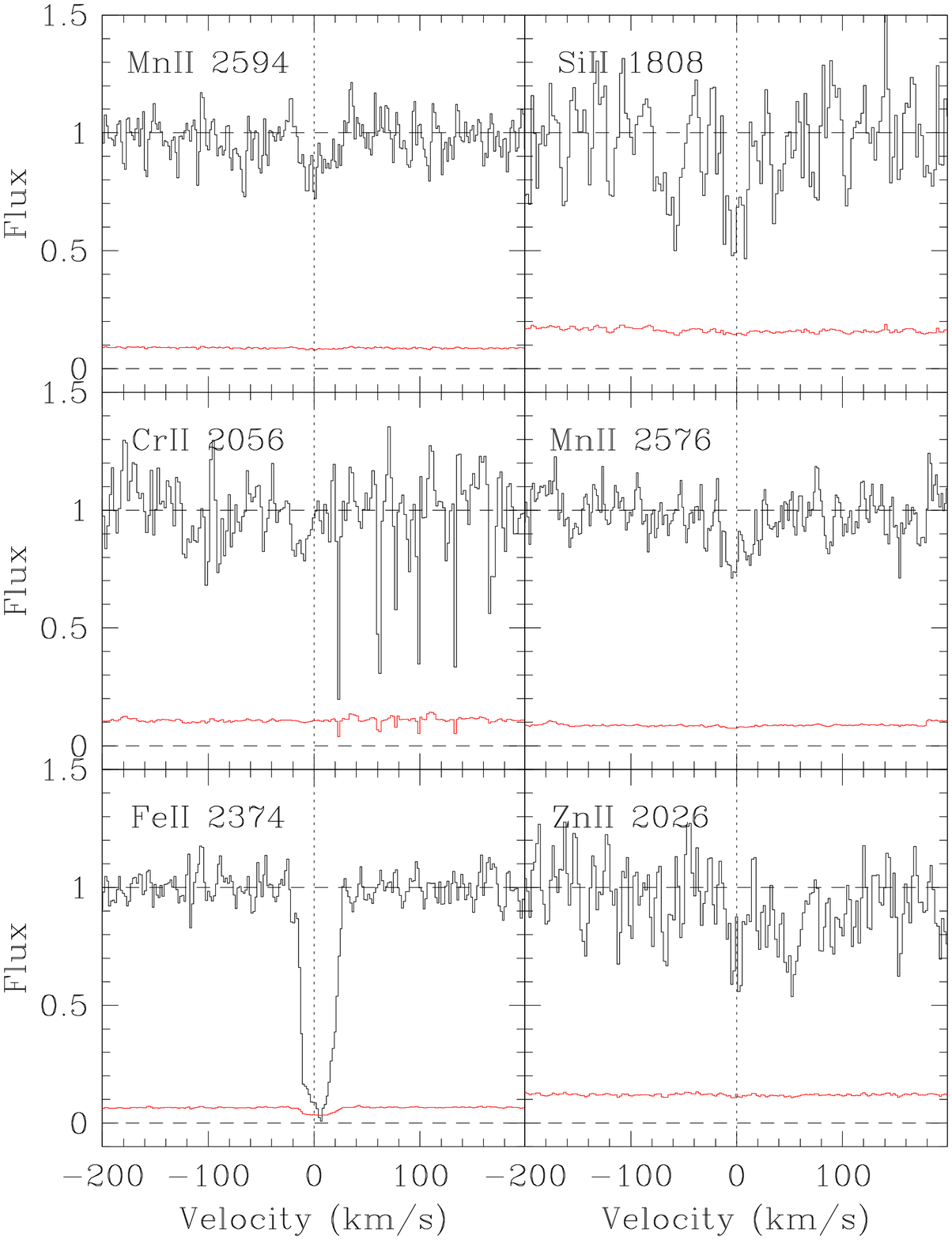}}}}
\caption{\label{B0227}Metal lines towards B0227$-$369 on a velocity scale 
relative to $z=1.02900$  }
\end{figure}

\begin{figure}
\centerline{\rotatebox{0}{\resizebox{7cm}{!}
{\includegraphics{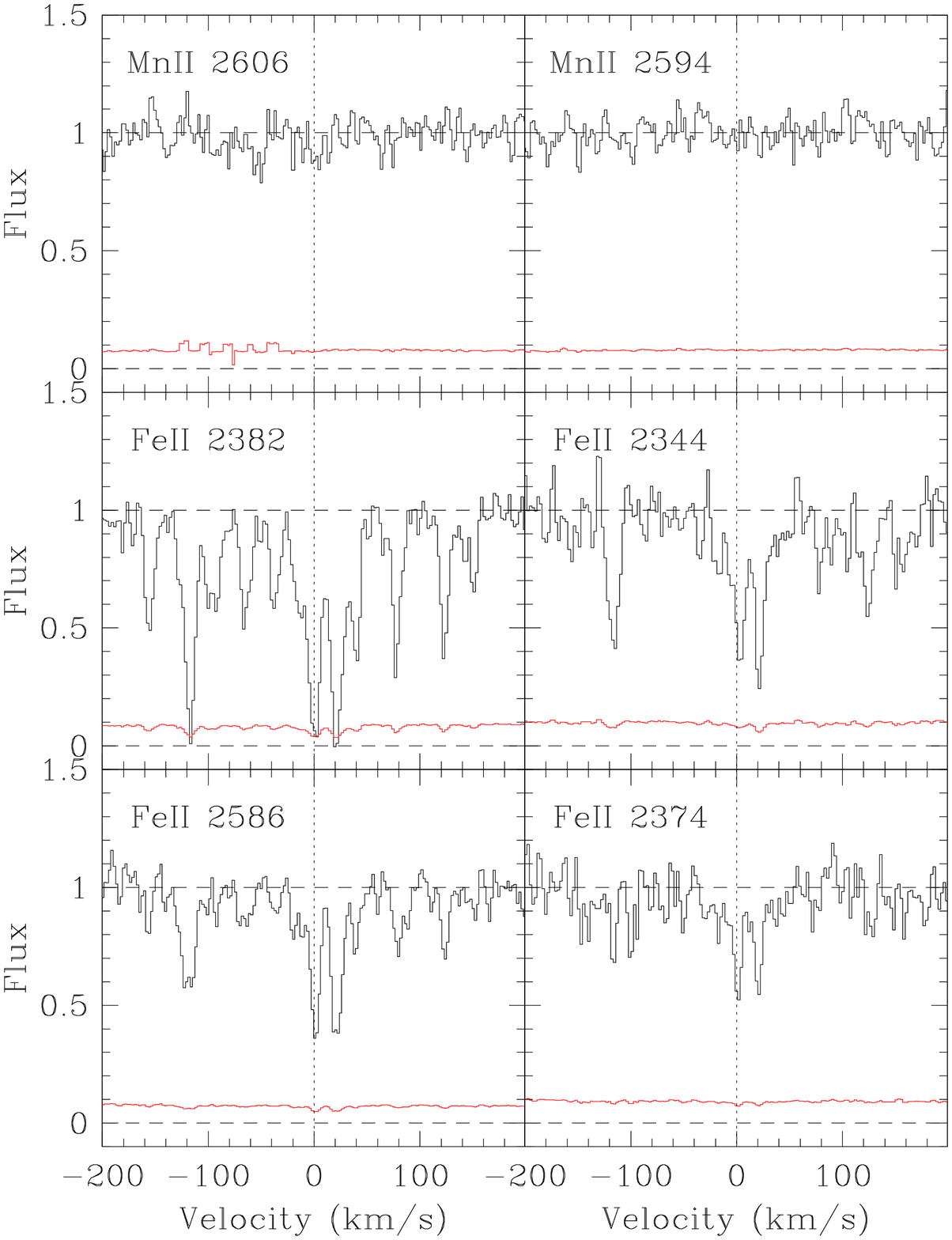}}}}
\caption{\label{B0240_1}Metal lines towards B0240$-$060 on a velocity scale 
relative to $z=0.58103$  }
\end{figure}

\begin{figure}
\centerline{\rotatebox{0}{\resizebox{7cm}{!}
{\includegraphics{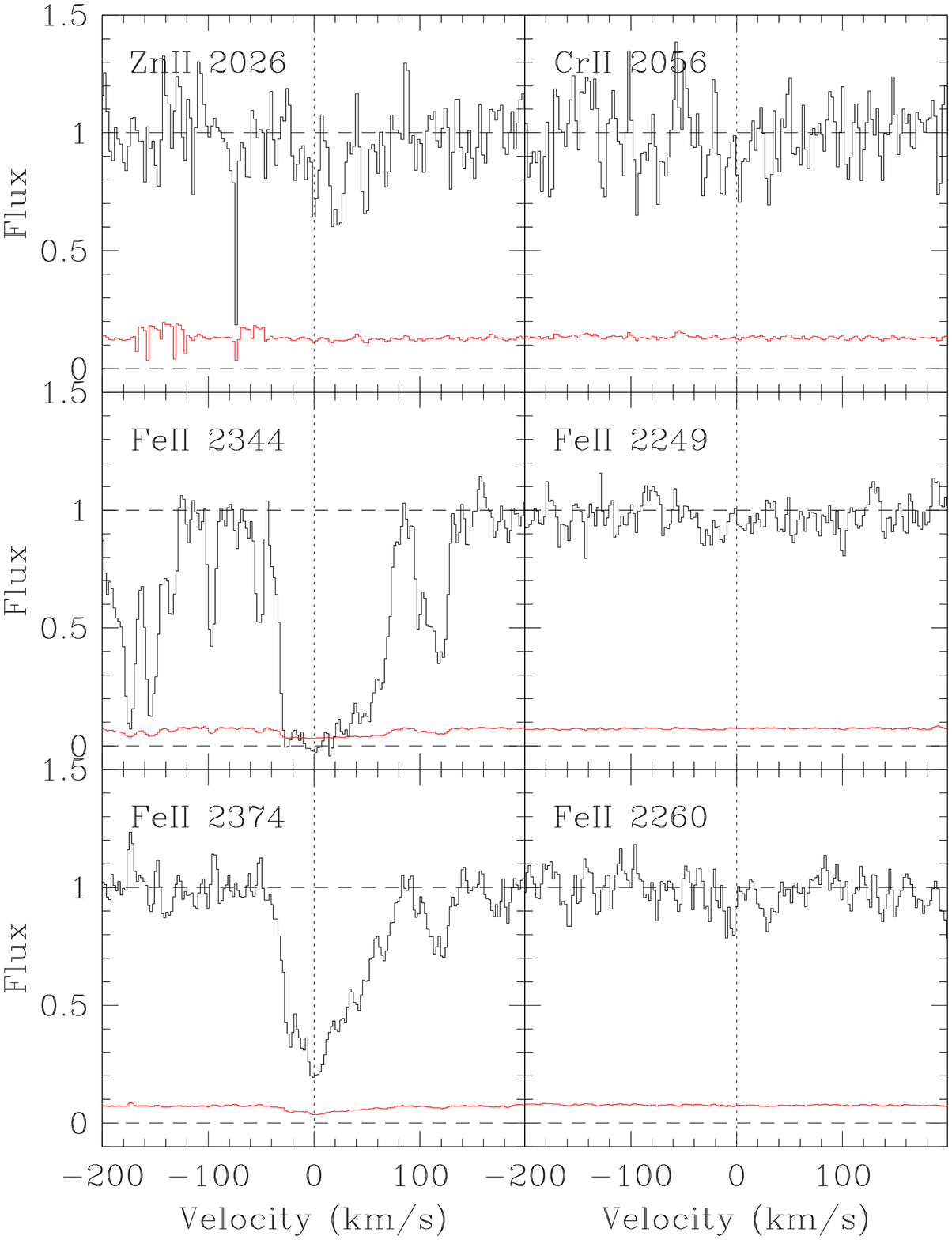}}}}
\caption{\label{B0240_2}Metal lines towards B0240$-$060 on a velocity scale 
relative to $z=0.75468$  }
\end{figure}

\begin{figure}
\centerline{\rotatebox{0}{\resizebox{7cm}{!}
{\includegraphics{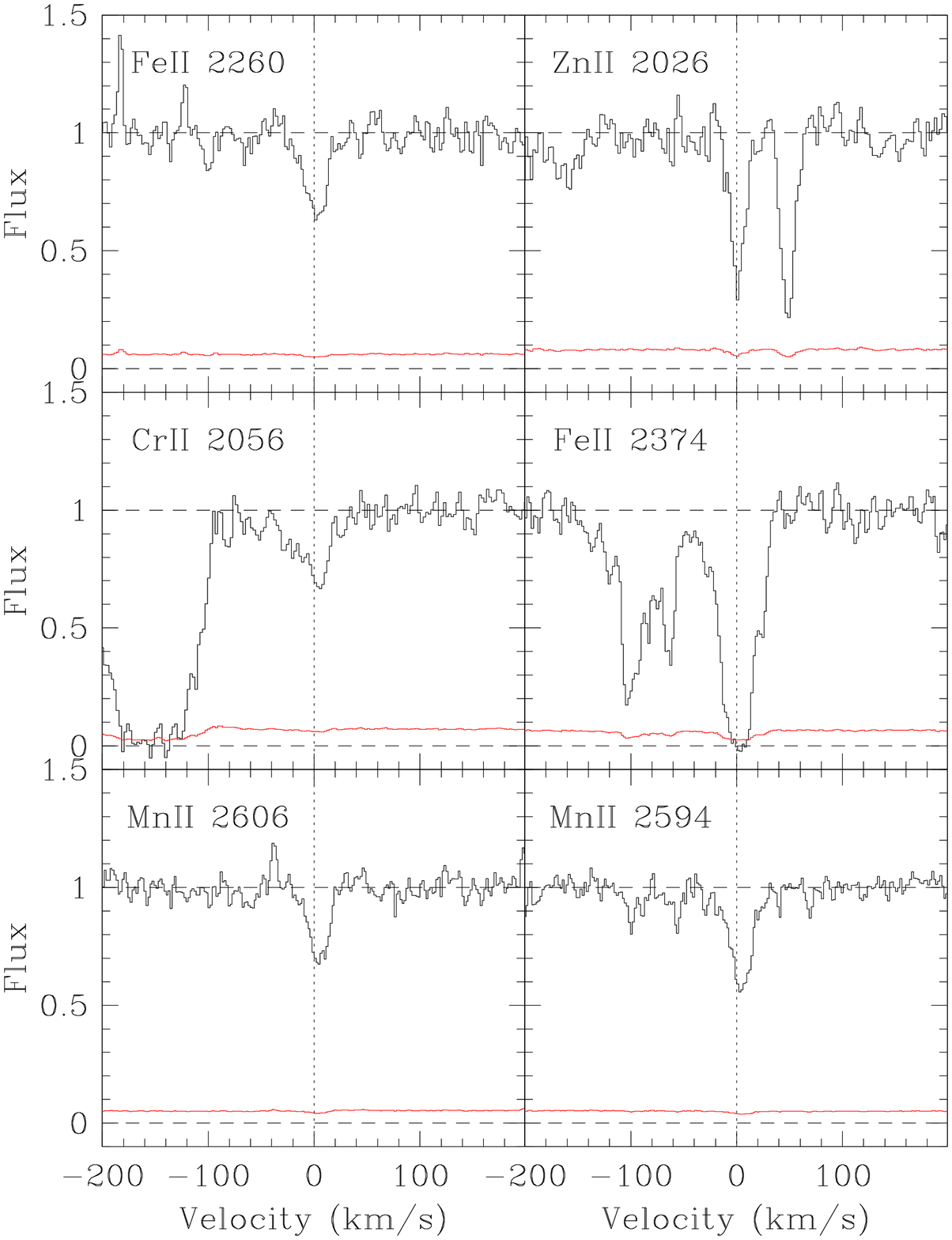}}}}
\caption{\label{B0244}Metal lines towards B0244$-$128 on a velocity scale 
relative to $z=0.82850$  }
\end{figure}

\begin{figure}
\centerline{\rotatebox{0}{\resizebox{7cm}{!}
{\includegraphics{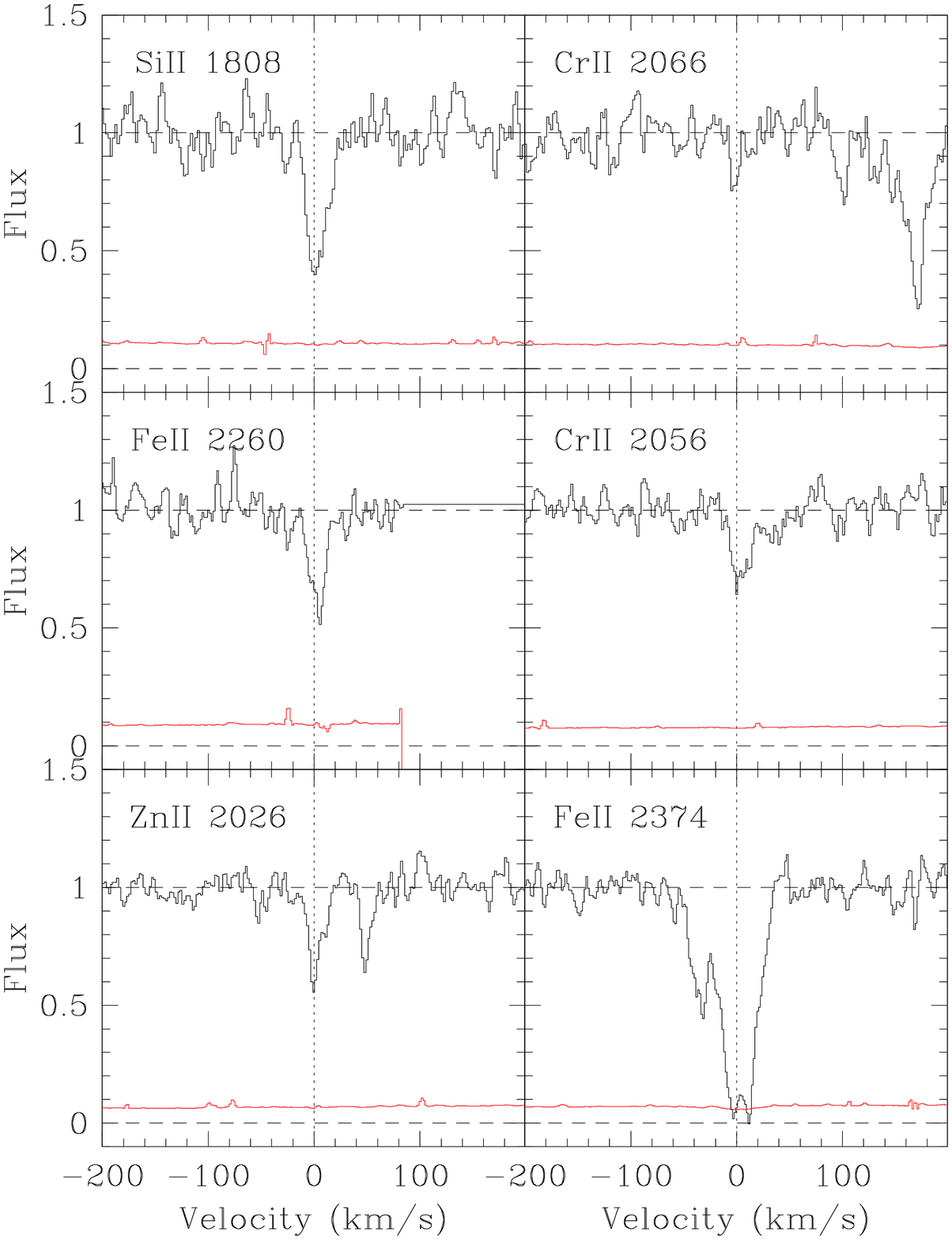}}}}
\caption{\label{B0458}Metal lines towards B0458$-$020 on a velocity scale 
relative to $z=1.56055$  }
\end{figure}

\begin{figure}
\centerline{\rotatebox{0}{\resizebox{7cm}{!}
{\includegraphics{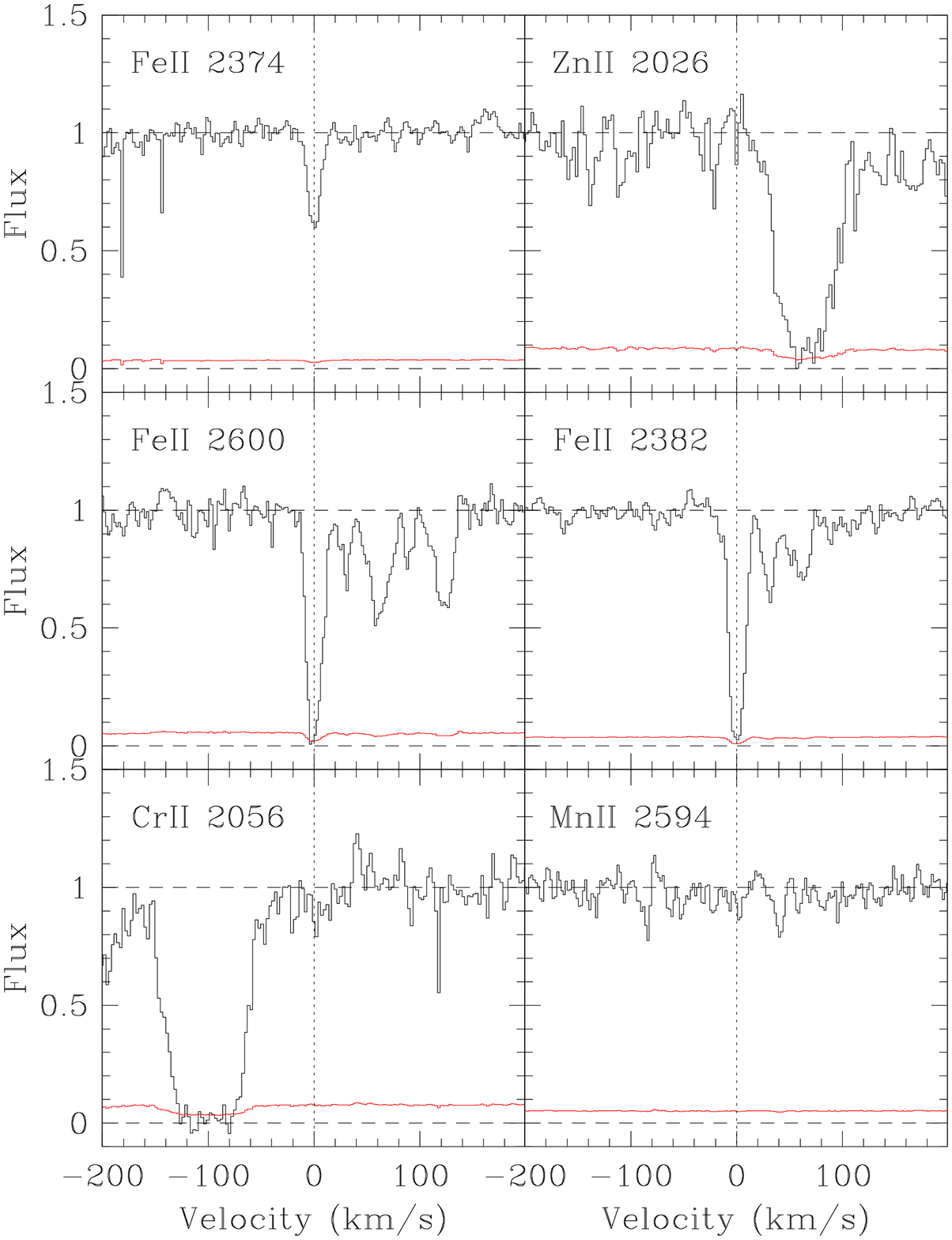}}}}
\caption{\label{B0919}Metal lines towards B0919$-$260 on a velocity scale 
relative to $z=0.70526$  }
\end{figure}

\begin{figure}
\centerline{\rotatebox{0}{\resizebox{7cm}{!}
{\includegraphics{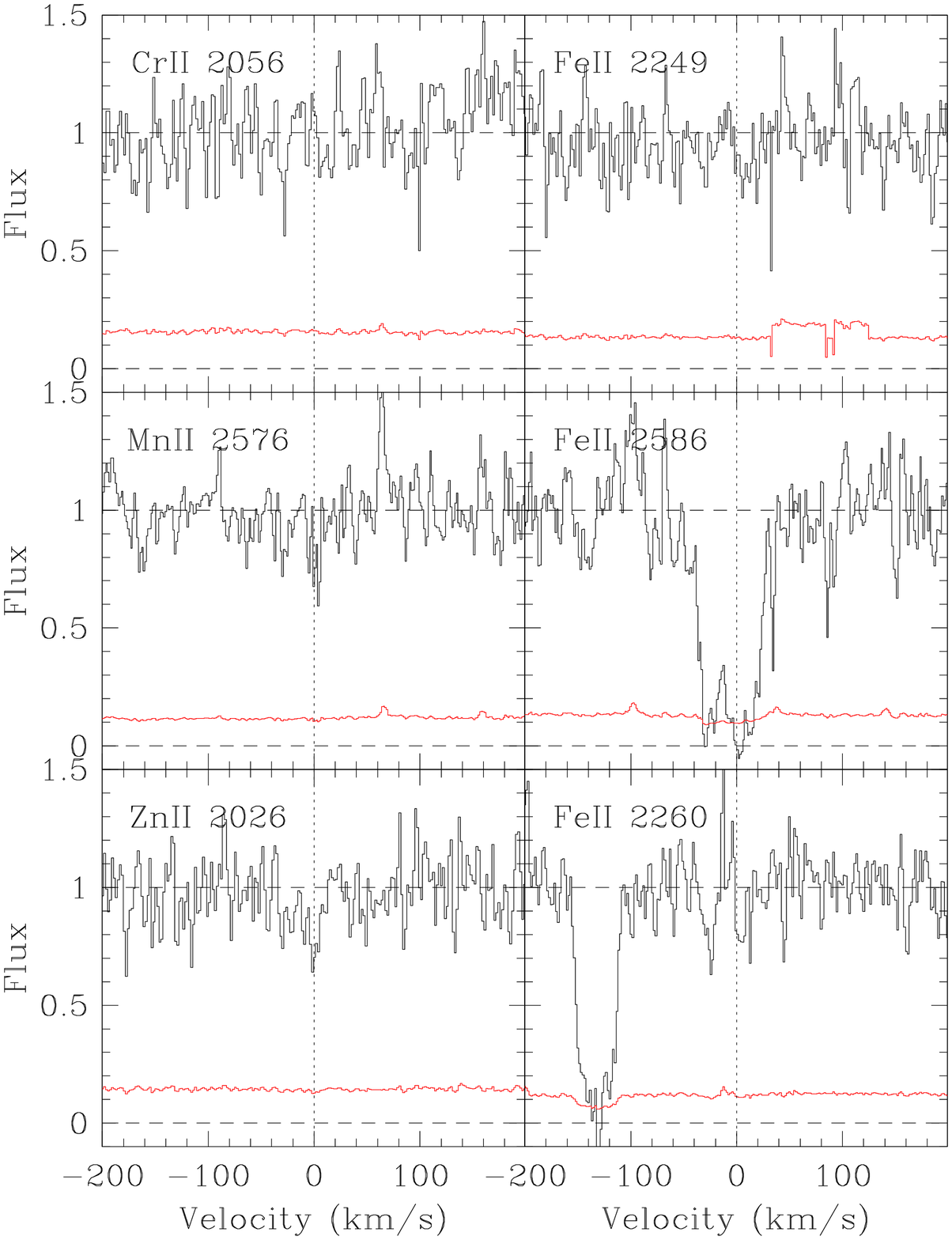}}}}
\caption{\label{B1005}Metal lines towards B1005$-$333 on a velocity scale 
relative to $z=1.37381$ }
\end{figure}

\begin{figure}
\centerline{\rotatebox{0}{\resizebox{7cm}{!}
{\includegraphics{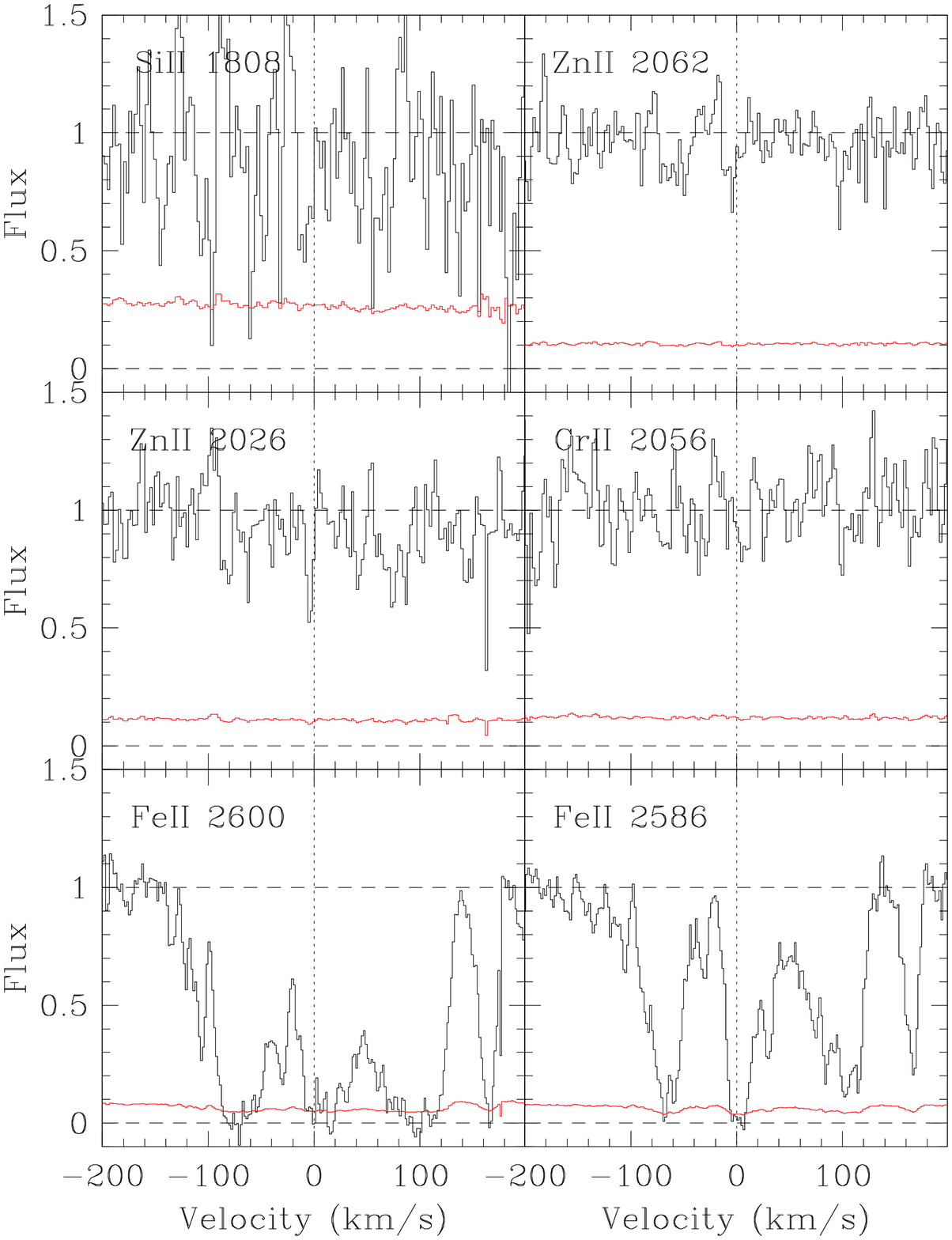}}}}
\caption{\label{B1256}Metal lines towards B1256$-$177 on a velocity scale 
relative to $z=0.93495$ }
\end{figure}

\begin{figure}
\centerline{\rotatebox{0}{\resizebox{7cm}{!}
{\includegraphics{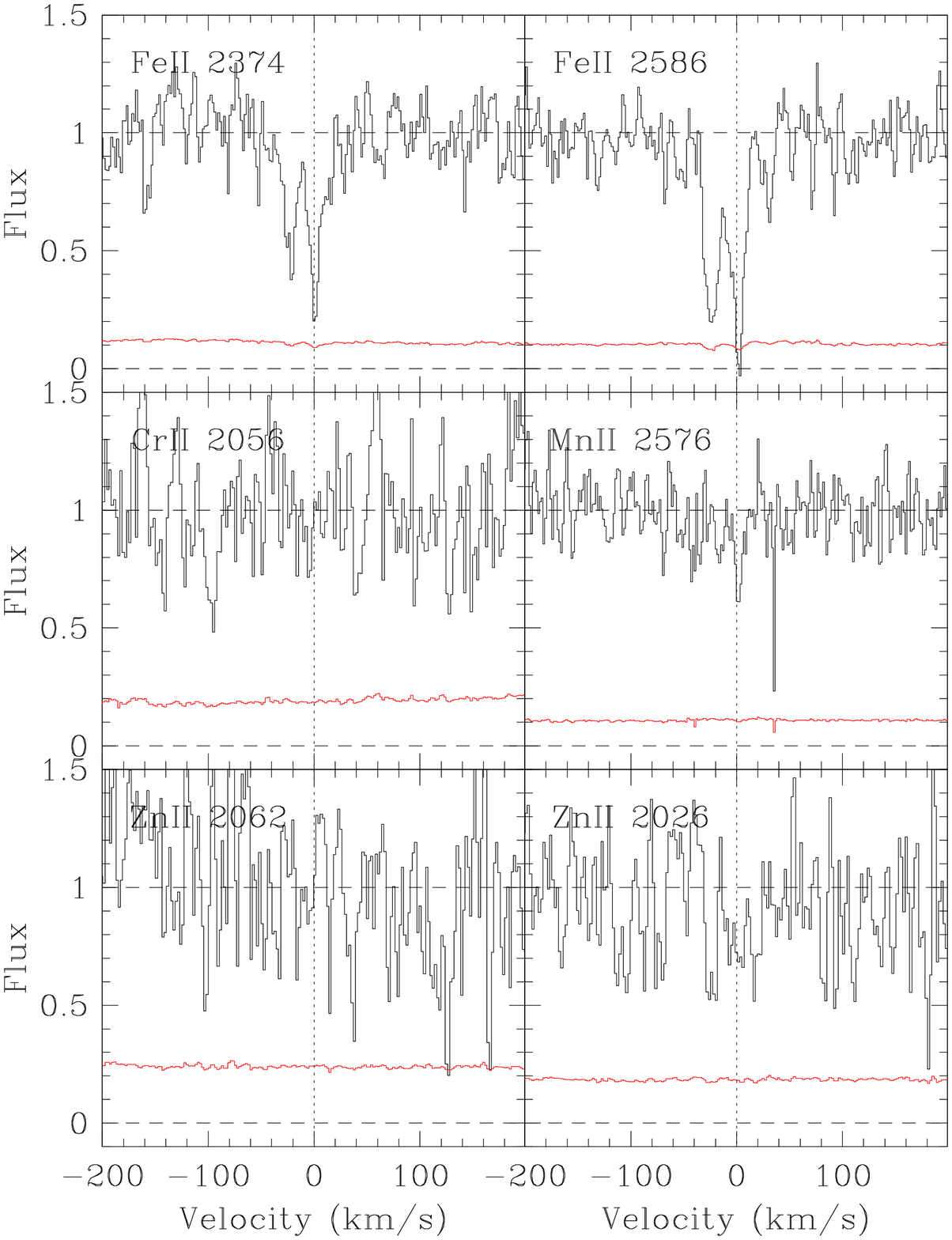}}}}
\caption{\label{B1318}Metal lines towards B1318$-$263 on a velocity scale 
relative to $z=1.10407$ }
\end{figure}

\begin{figure}
\centerline{\rotatebox{0}{\resizebox{7cm}{!}
{\includegraphics{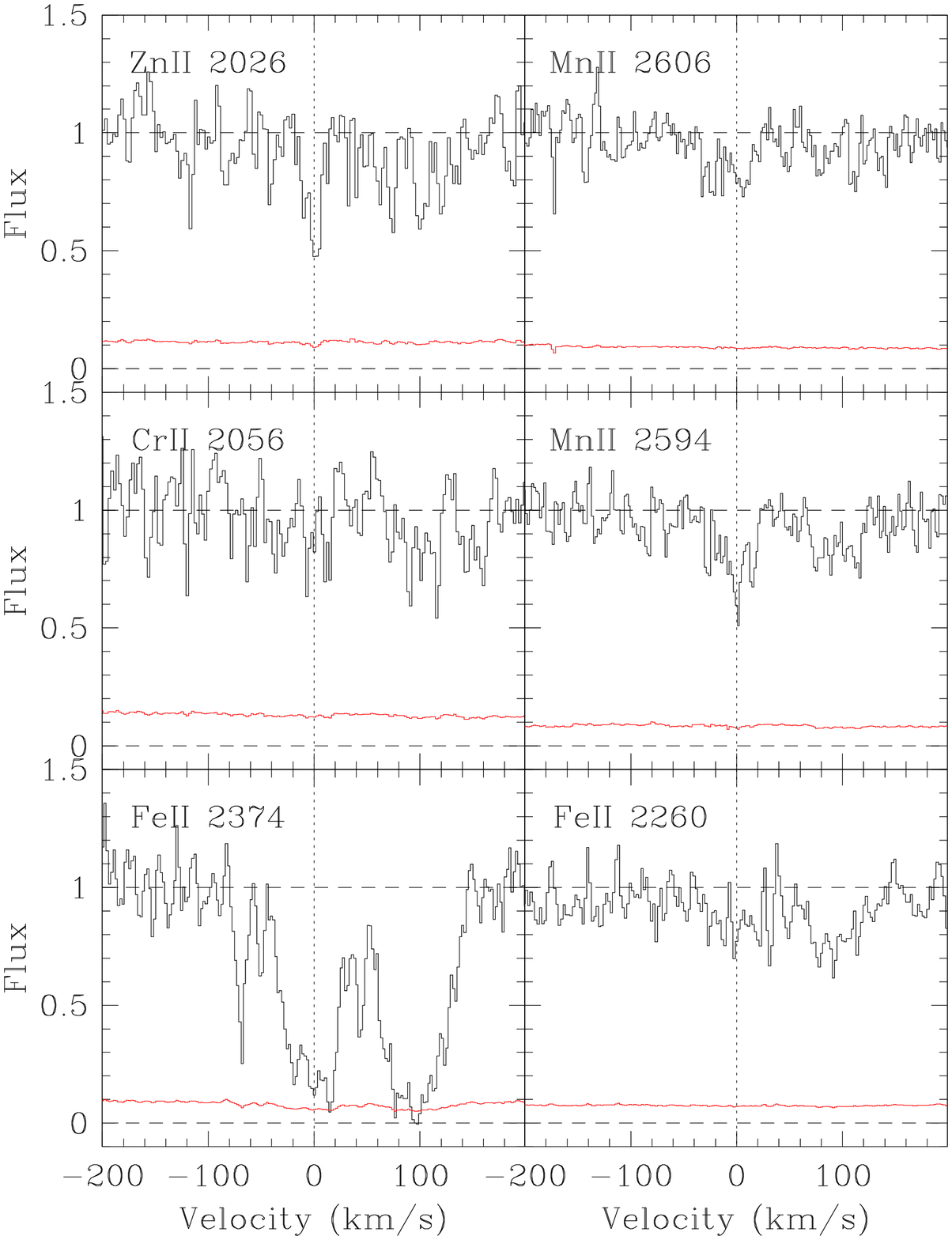}}}}
\caption{\label{B1324}Metal lines towards B1324$-$047 on a velocity scale 
relative to $z=0.78472$ }
\end{figure}

\begin{figure}
\centerline{\rotatebox{0}{\resizebox{7cm}{!}
{\includegraphics{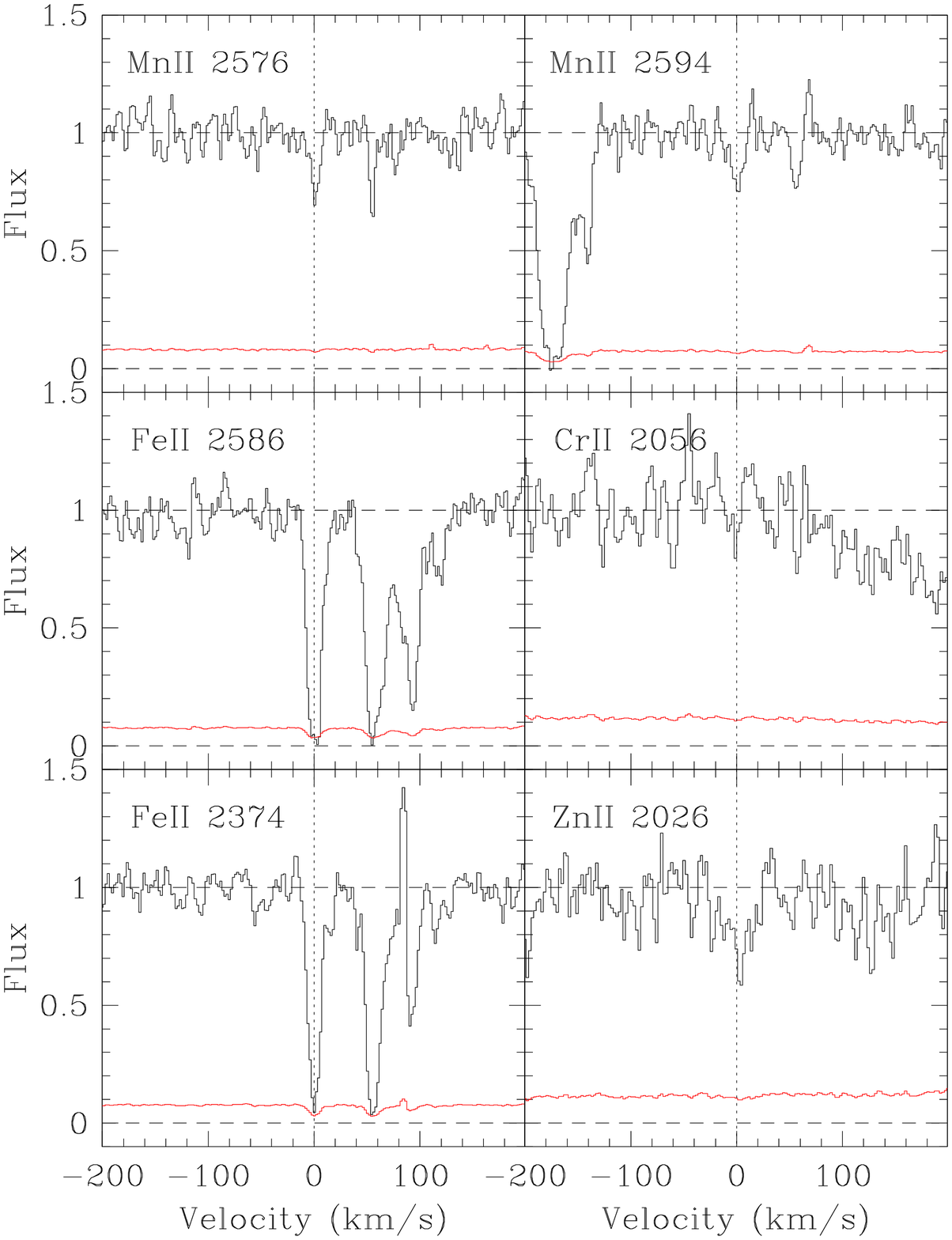}}}}
\caption{\label{B1402}Metal lines towards B1402$-$012  on a velocity scale 
relative to $z=0.88978$  }
\end{figure}

\begin{figure}
\centerline{\rotatebox{0}{\resizebox{7cm}{!}
{\includegraphics{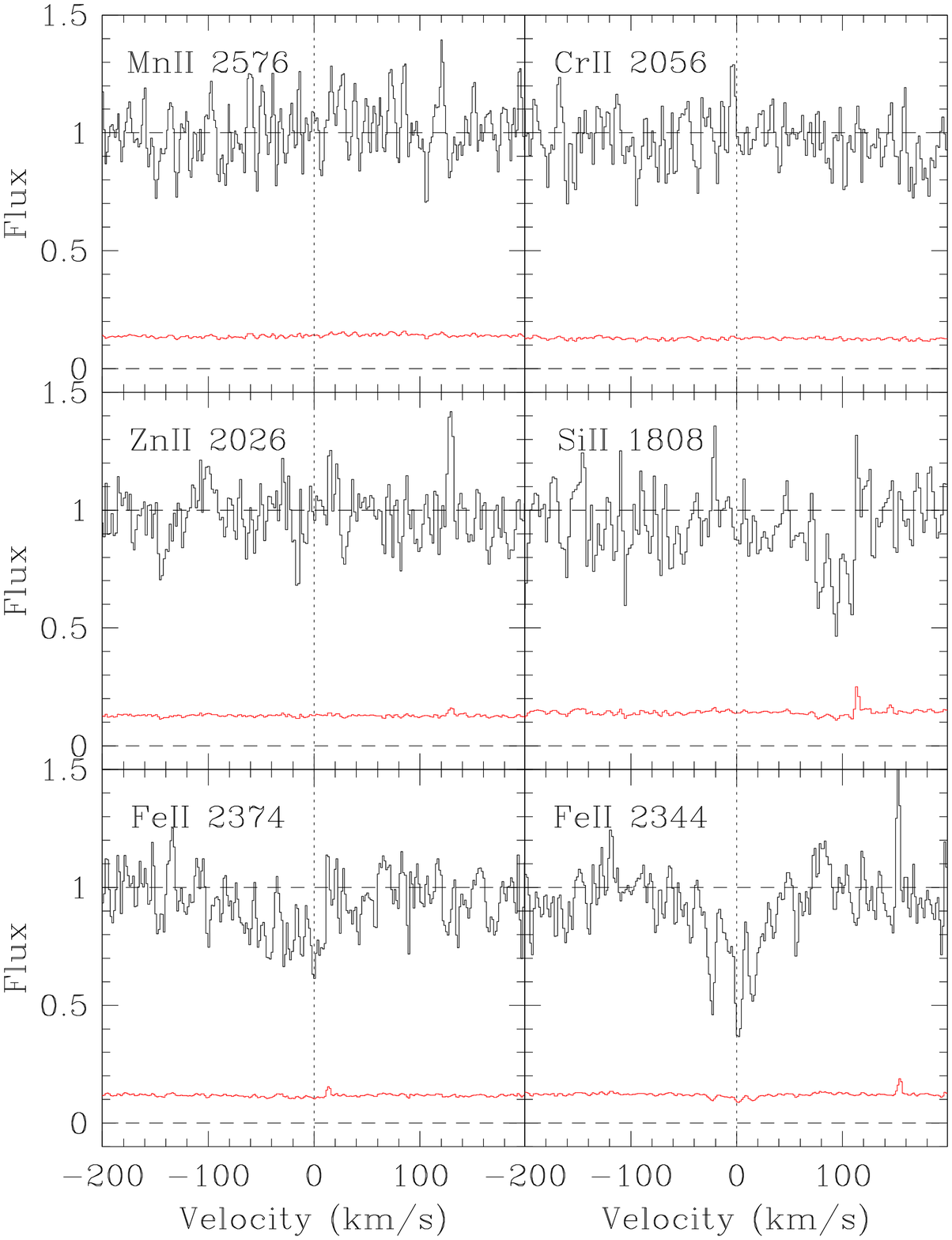}}}}
\caption{\label{B1412}Metal lines towards B1412$-$096  on a velocity scale 
relative to $z=1.34652$  }
\end{figure}

\begin{figure}
\centerline{\rotatebox{0}{\resizebox{7cm}{!}
{\includegraphics{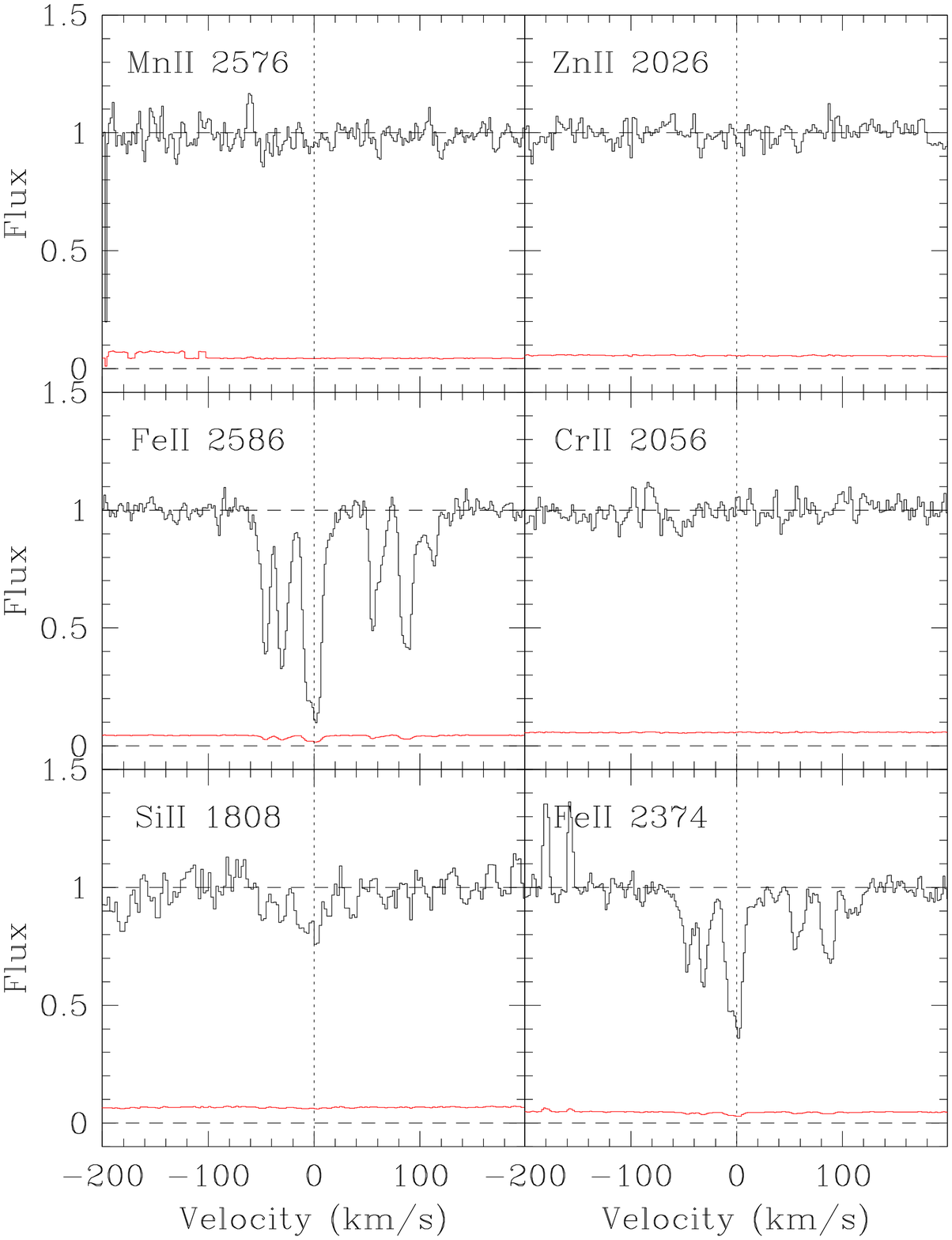}}}}
\caption{\label{B2149_1}Metal lines towards B2149$-$307 on a velocity scale 
relative to $z=1.09074$ }
\end{figure}

\begin{figure}
\centerline{\rotatebox{0}{\resizebox{7cm}{!}
{\includegraphics{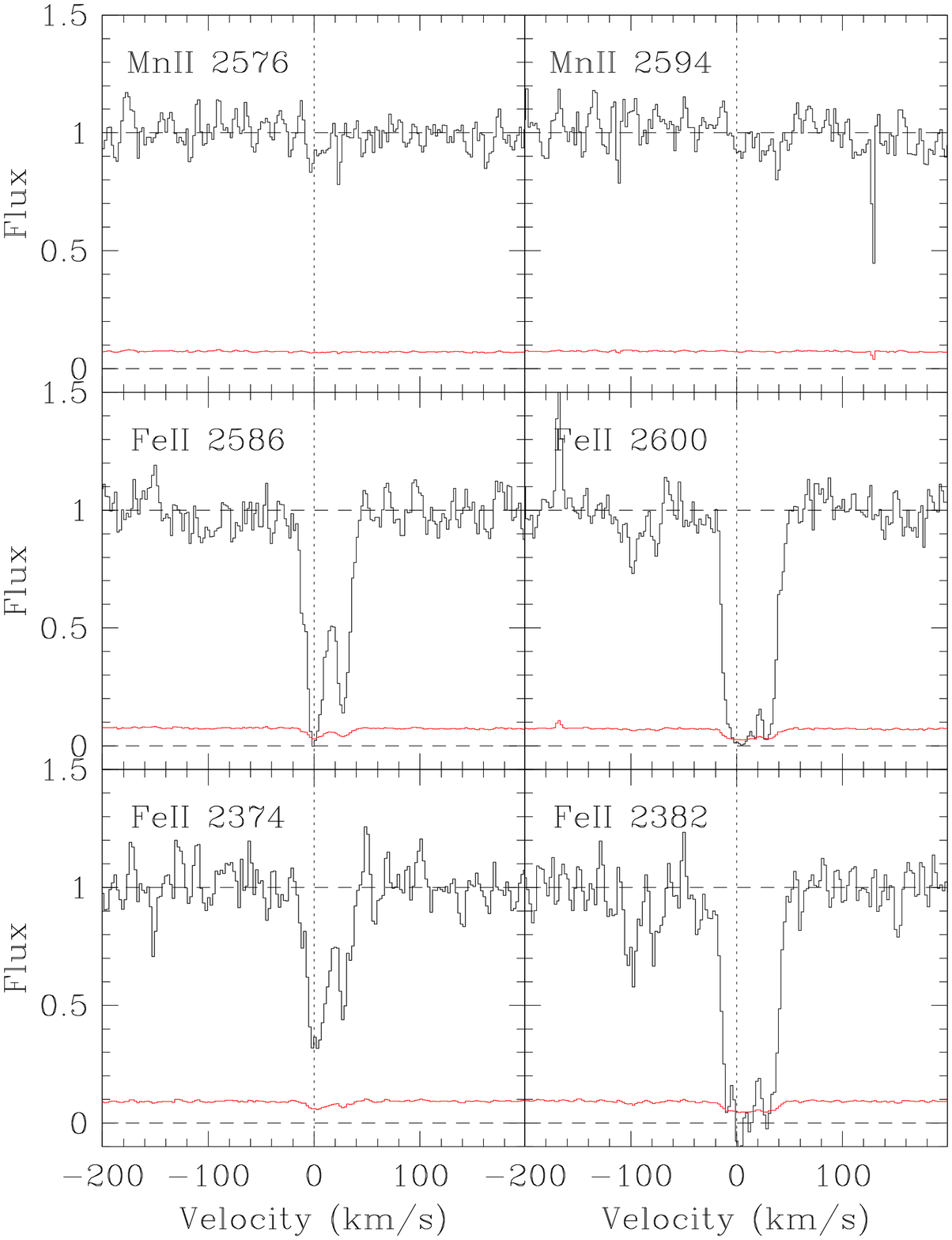}}}}
\caption{\label{B2245} Metal lines towards B2245$-$128 on a velocity scale 
relative to $z=0.58700$}
\end{figure}

\begin{figure}
\centerline{\rotatebox{0}{\resizebox{7cm}{!}
{\includegraphics{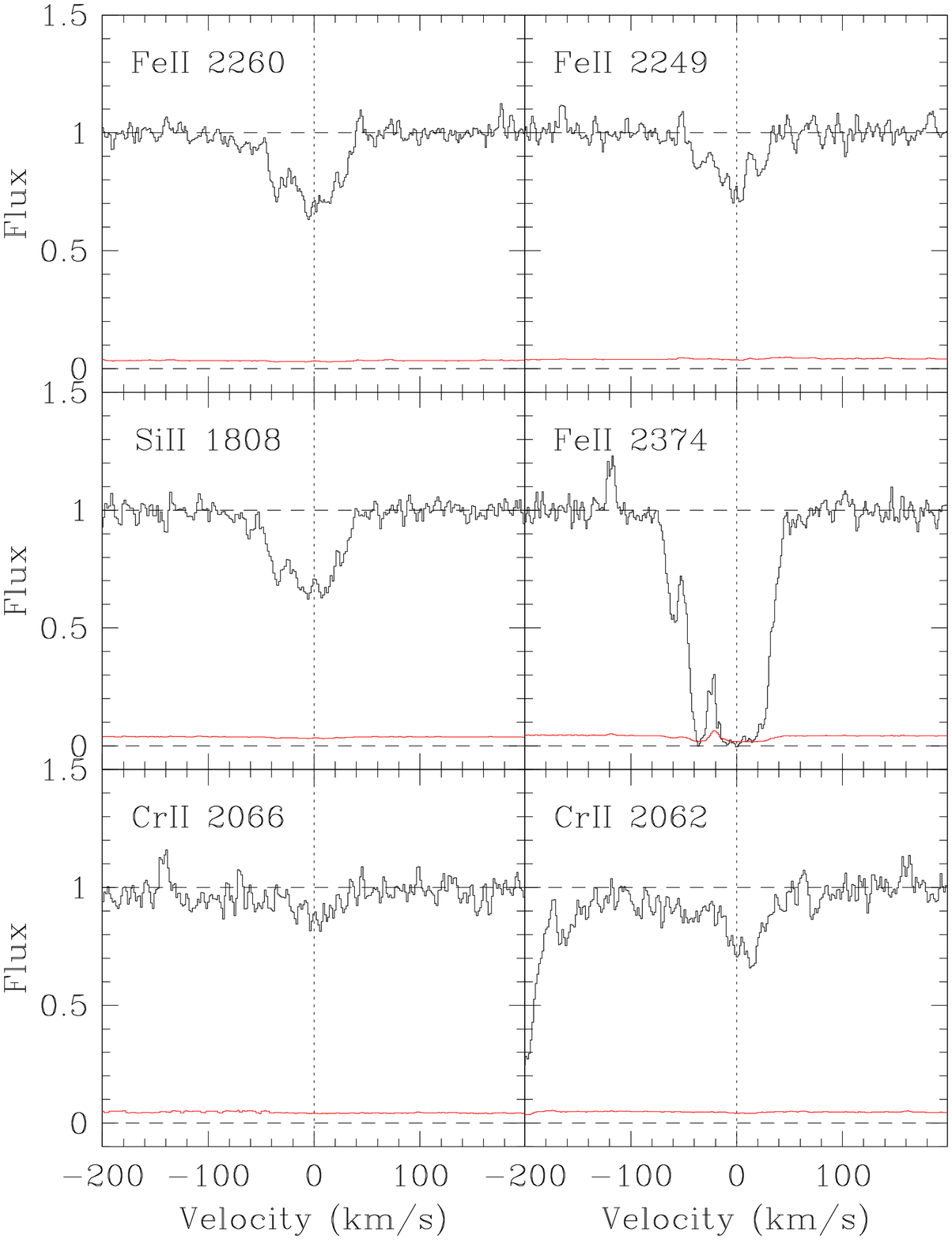}}}}
\caption{\label{B0122_1}Metal lines towards B0122$-$005 on a velocity scale 
relative to $z=1.76085$  }
\end{figure}

\begin{figure}
\centerline{\rotatebox{0}{\resizebox{7cm}{!}
{\includegraphics{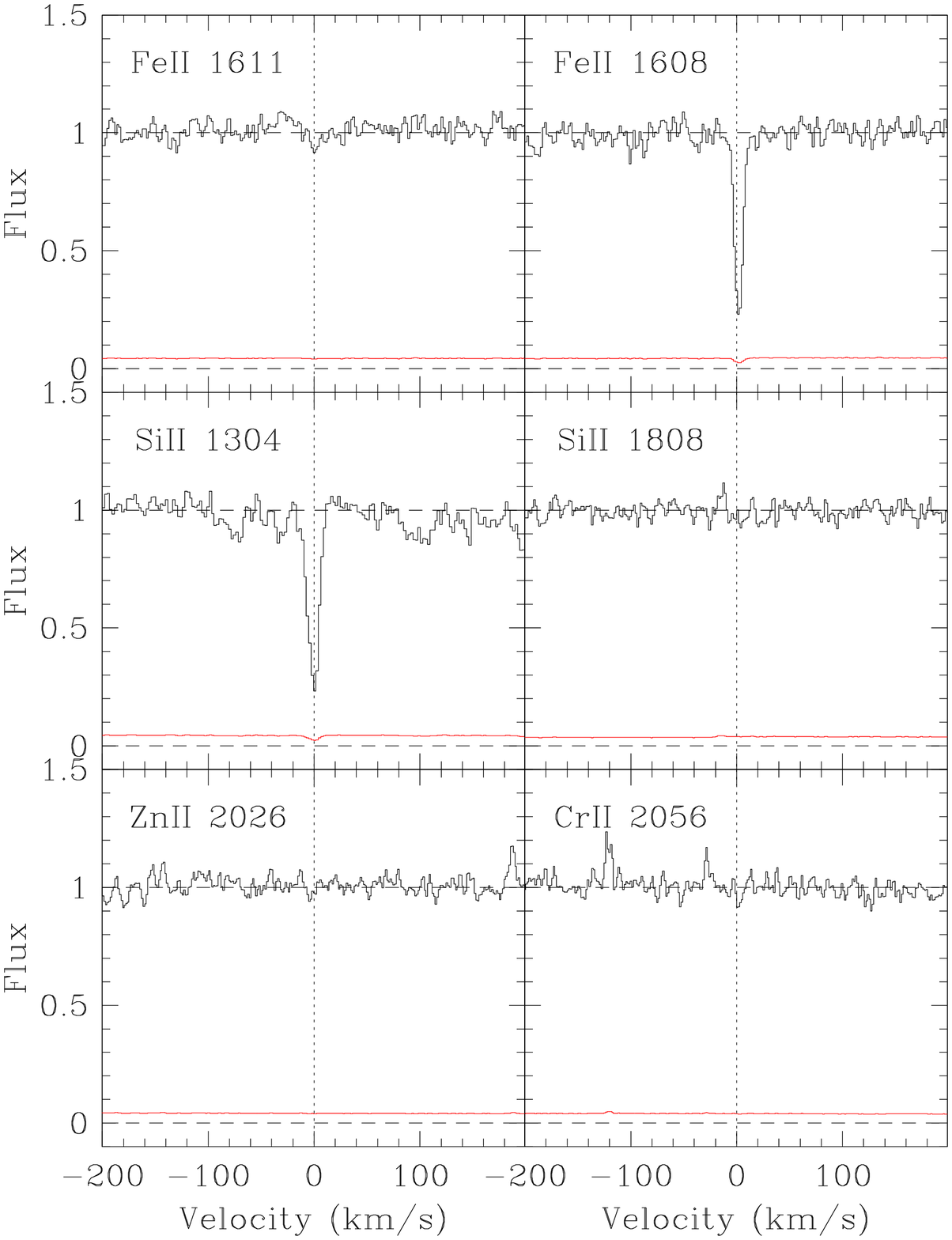}}}}
\caption{\label{B0122_2}Metal lines towards B0122$-$005 on a velocity scale 
relative to $z=2.00950$  }
\end{figure}

\begin{figure}
\centerline{\rotatebox{0}{\resizebox{7cm}{!}
{\includegraphics{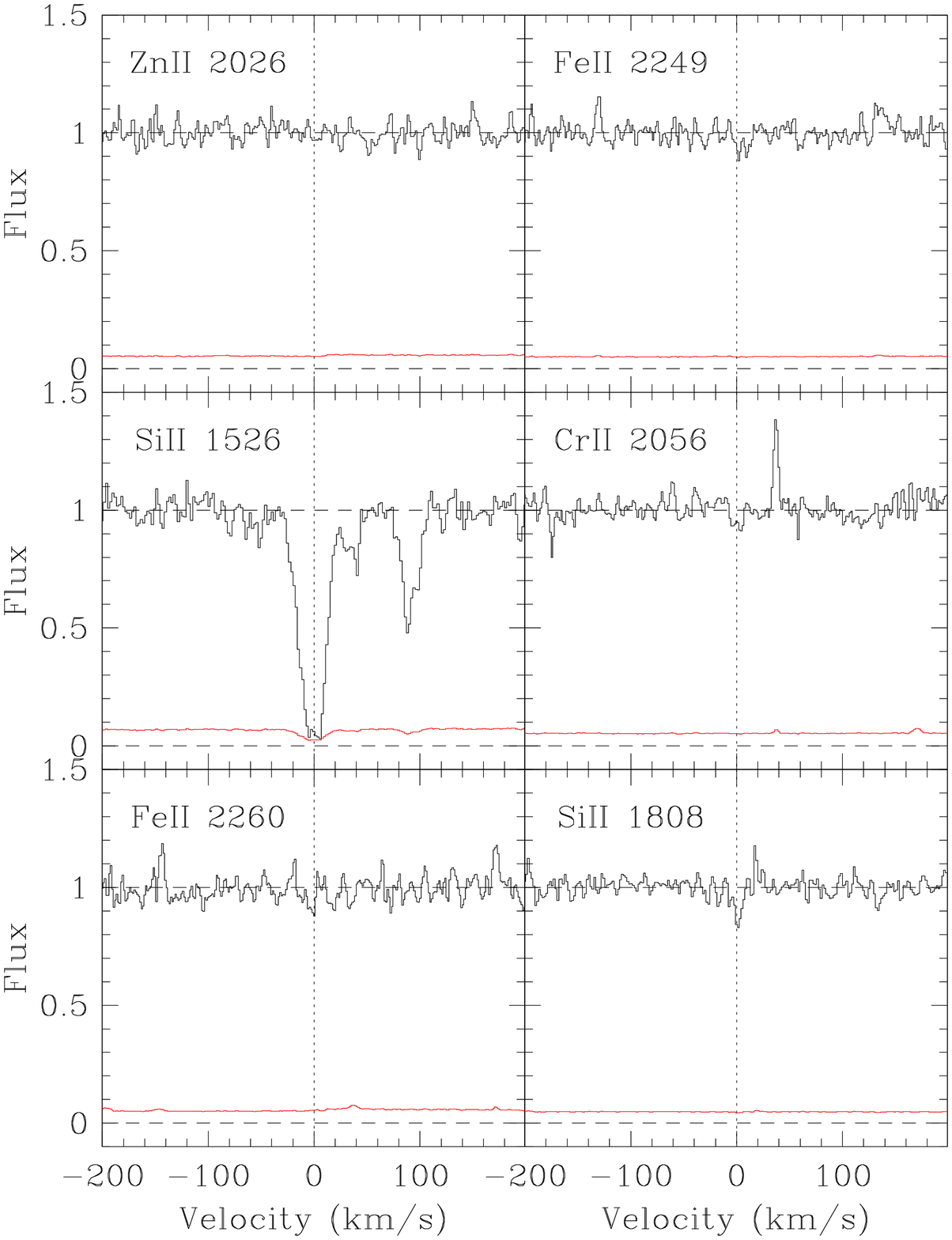}}}}
\caption{\label{B0244_2}Metal lines towards B0244$-$128 on a velocity scale 
relative to $z=1.86278$  }
\end{figure}

\begin{figure}
\centerline{\rotatebox{0}{\resizebox{7cm}{!}
{\includegraphics{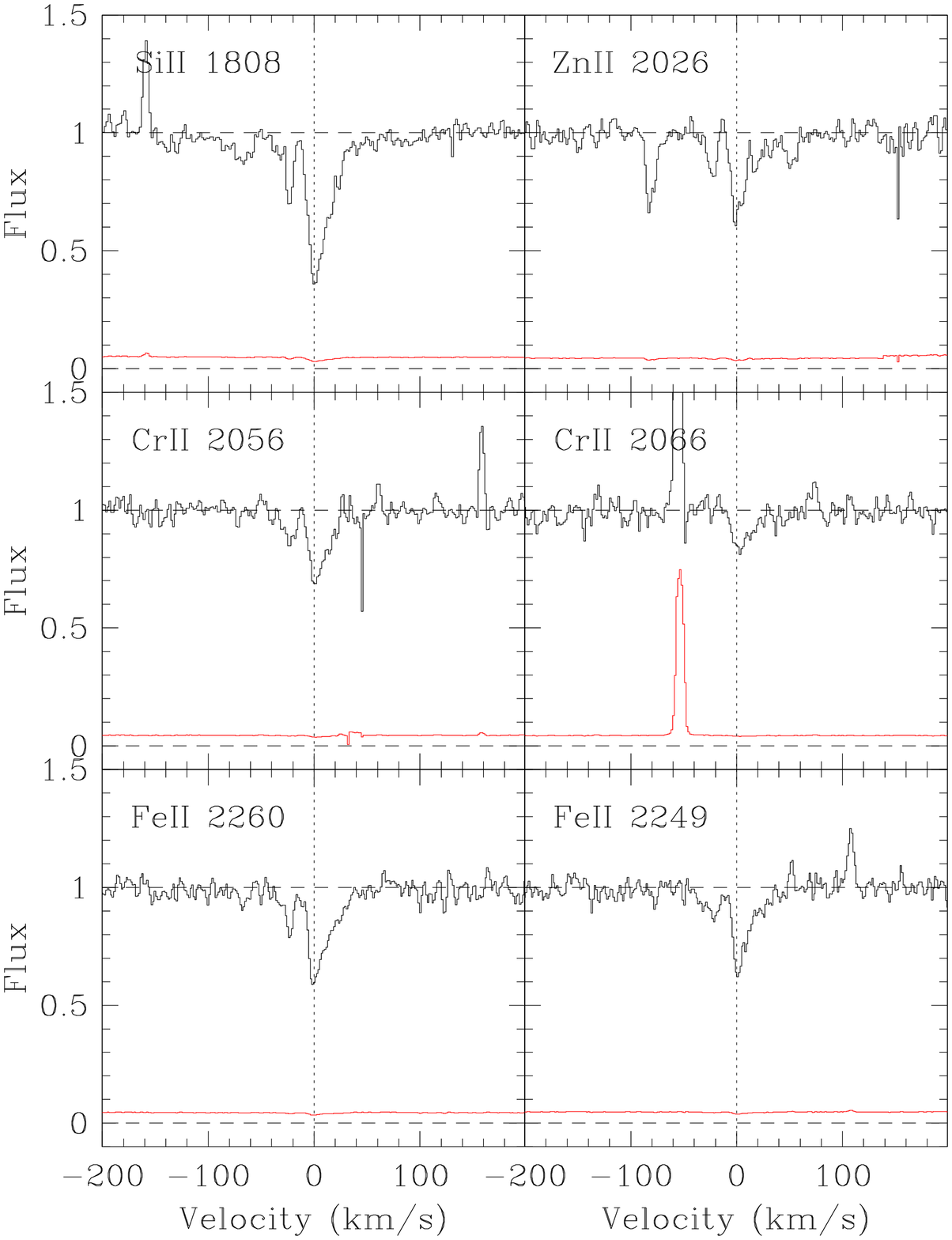}}}}
\caption{\label{B2149_2}Metal lines towards B2149$-$307 on a velocity scale 
relative to $z=1.70085$ }
\end{figure}

Figures \ref{B0039} -- \ref{B2149_2} show selected metal lines
for all absorption systems included in this study.

% New DLAs

\clearpage

\clearpage

\clearpage

\section{Atomic data}

In Table \ref{atom_data} we reproduce the atmoic data used in our column density
determinations.

\begin{center}
\begin{table}
\caption{\label{atom_data}Atomic data for the detected transitions from Morton (2003)}
\begin{tabular}{lcc}
\hline
Species & $\lambda$ & $f$ \\ \hline
\si2 & 1304.3702 & 0.094 \\
\si2 & 1808.0130 & 0.002186 \\
\chr2 & 2056.2539 & 0.10500 \\ 
\chr2 & 2062.2340 & 0.07800 \\ 
\chr2 & 2066.1610 & 0.05150 \\ 
\fe2 & 2600.1729 & 0.2390   \\ 
\fe2 & 2586.6500 & 0.06910 \\
\fe2 & 2382.7650 & 0.3200  \\
\fe2 & 2374.4612 & 0.0313 \\
\fe2 & 2344.2140 & 0.1140 \\
\fe2 & 2260.7805 & 0.00244\\
\fe2 & 2249.8768 & 0.001821 \\
\fe2 & 1608.4511 & 0.05800 \\
\zn2 & 2062.6640 & 0.2560 \\
\zn2 & 2026.1360 & 0.4890 \\
\mn2 & 2606.4620 & 0.1927 \\
\mn2 & 2594.4990 & 0.2710 \\
\mn2 & 2576.8770 & 0.3508 \\
 \hline 
\end{tabular}
\end{table}
\end{center}

\end{appendix}

\end{document}